\documentclass[aps,prc,twocolumn,superscriptaddress,preprintnumbers,amsmath,amssymb,showkeys,floatfix,nofootinbib,reprint]{revtex4-1}

\usepackage{amsfonts}
\usepackage{amssymb}
\usepackage{graphics}
\usepackage{graphicx}
\usepackage{epstopdf}
\usepackage{dcolumn}
\usepackage{bm}
\usepackage{longtable}
\usepackage{epsfig}
\usepackage{times}
\usepackage{url}
\usepackage{color}

\begin{document}

\title{Efficient production of $^{229m}$Th via nuclear excitation by electron capture}

%
%

\author{Jingyan \surname{Zhao}}
\affiliation{School of Physics, Nankai University, Tianjin 300071, China}

\author{Adriana \surname{P\'alffy}}
\affiliation{University of W\"urzburg, Institute of Theoretical Physics and Astrophysics, Am Hubland, 97074 W\"urzburg, Germany}

\author{Christoph H. \surname{Keitel}}
\affiliation{Max-Planck-Institut f\"ur Kernphysik, Saupfercheckweg 1, D-69117 Heidelberg, Germany}

\author{Yuanbin \surname{Wu}}
\email{yuanbin@nankai.edu.cn}
\affiliation{School of Physics, Nankai University, Tianjin 300071, China}

\date{\today}

\begin{abstract}

The nuclear isomeric state $^{229m}$Th with an exceptionally low excitation energy makes the $^{229}$Th isotope a crucial candidate for nuclear clocks and many other applications. Efficient and controllable production of $^{229m}$Th is essential and still remains a challenge. Here we report a novel approach for efficient production of $^{229m}$Th by the excitation of $^{229}$Th to the above-lying excited state at $29.19$ keV energy via the process of nuclear excitation by electron capture (NEEC). We show theoretically that the production rate of $^{229m}$Th per nucleus with accessible conditions can be six orders of magnitude larger than the value experimentally demonstrated using $29$-keV synchrotron radiation for this indirect excitation. With the efficient production of $^{229m}$Th, our results identify scenarios, as well as the characteristic NEEC signature with which NEEC events could be unambiguously identified, for a clear experimental identification of the long-sought NEEC phenomenon.

\end{abstract}

\maketitle


\section{Introduction}

The first nuclear excited state of the $^{229}$Th isotope, denoted here as $^{229m}$Th, is the lowest known metastable nuclear excited state \cite{Walker1999, Walker2020} at an energy about $8$ eV above the nuclear ground state \cite{PeikEPL2003,WenseEPJA2020,BeeksNRP2021}. As this extremely low excitation energy could enable direct laser manipulation of nuclear states, the $^{229}$Th isotope has attracted a great deal of attention as a possible nuclear clock which may exceed the present accuracy of atomic optical clocks \cite{PeikEPL2003,WenseEPJA2020,BeeksNRP2021,CampbellPRL2012,KazakovNJP2012,PeikQST2021}. Furthermore, this unique isomer, owing to its extremely low excitation energy and potential for a nuclear frequency standard, opens also opportunities for many prospects, including the studies of fundamental interactions \cite{FlambaumPRL2006,FlambaumPRL2016}, the detection of dark matter \cite{PeikQST2021,FadeevPRA2020}, and interdisciplinary applications \cite{BeeksNRP2021,TkalyaPRL2011,ThirolfAP2019}.

A major focus is the efficient and controllable production of $^{229m}$Th, as well as the precise measurement of properties of $^{229m}$Th. The $\alpha$-decay of $^{233}$U \cite{WenseNat2016,ThielkingNat2018,SeiferlePRL2017,SeiferleNat2019,YamaguchiPRL2019,SikorskyPRL2020,YamaguchiNat2024} or the $\beta$-decay of $^{229}$Ac \cite{KraemerNat2023} to populate the isomeric state and measure its properties such as the excitation energy and lifetime have been demonstrated experimentally. However, nuclear decays may not be controllable as they are statistical phenomena. An alternative method which has been demonstrated experimentally for the production of $^{229m}$Th is the indirect excitation via x-ray pumping of the second-excited state at the excitation energy of $29.19$ keV \cite{MasudaNat2019}: the second-excited state is populated by synchrotron radiation and partially decays to $^{229m}$Th. Very recently, direct laser excitation to $^{229m}$Th in Th-doped VUV-transparent crystals has also been demonstrated \cite{TiedauPRL2024,ElwellPRL2024}. Despite the fact that ultimately, the nuclear clock will require narrow-band laser driven excitation of the isomer \cite{WensePRL2017,KirschbaumPRC2022,JinPRR2023,TkalyaNPA2022,TiedauPRL2024,ElwellPRL2024}, several alternative excitation methods have been discussed in the literature, for instance, electron collisions and laser-cluster interaction \cite{TkalyaPRL2020,WangPRL2021,QiPRL2023}, the electron-bridge process \cite{PorsevPRA2010,NickersonPRL2020,BilousPRL2020,DzyublikPRC2022}, or other processes with coupling to electrons \cite{KarpeshinPLB1996,KarpeshinPRC2017,XuPRA2023}. Most of these methods take advantages of interactions between the nuclear and electronic degrees of freedom of the atom.

Driving nuclear transitions via coupling to the atomic shell is one of the most intriguing ways for the manipulation of nuclear states. In the resonant process of nuclear excitation by electron capture (NEEC), an electron recombines into an atomic vacancy of an ion with the simultaneous excitation of the nucleus \cite{GoldanskiiPLB1976,PalffyCP2010}. The NEEC process was originally predicted in 1976 \cite{GoldanskiiPLB1976}, and it is the time-reversed internal conversion (IC). In this process, the transition energy of the electron recombination has to match the nuclear transition energy. NEEC has been investigated theoretically or envisaged in channeling through crystals \cite{Cue1989,Kimball1,Kimball2}, laser-generated or astrophysical-type plasmas \cite{HarstonPRC1999,GosselinPRC2004,GosselinPRC2007,GunstPRL2014,GunstPOP2015,WuPRL2018,QiPRL2023,SpohrEPJA2023,NegoitaRRP2016,NIF2018,doolen1978nuclear,doolen1978nuclearC}, storage rings \cite{PalffyPRA2006,PalffyPLB2008,MaPS2015,NIF2018,YangFP2024}, or electron beam ion traps (EBITs) \cite{PalffyPRA2007,DillmannBerlin2022,WangFP2023}. NEEC with reshaping of electron wave functions \cite{Madan2020,WuPRL2022} or with excited ions \cite{GargiuloPRL2022} has also been studied theoretically, with the results showing interesting features. In the context of $^{229}$Th, NEEC was investigated theoretically for the scenario of the direct excitation to $^{229m}$Th in the plasma generated in the laser-cluster interaction \cite{QiPRL2023}. The first claim of experimental observation for NEEC was reported recently in the $^{93m}$Mo isomer depletion in a beam-based scenario \cite{ChiaraNature2018}. However, subsequent theoretical calculations contradict the experimental value \cite{Wu2019,Polasik2021,RzadkiewiczPRC2023}, and the subject became one of the current focused debates in the field of nuclear physics \cite{Guo2021,Chiara2021}. A new and similar experiment has been performed with an isomer beam, but no signal of isomer depletion was observed \cite{GuoPRL2022}. Further investigations and conclusive observations on NEEC are highly demanded. Clean environments which could lead to a control of the NEEC process with NEEC being the dominating nuclear excitation mechanism should be helpful. Identifying the imprint of NEEC which could distinguish NEEC events from other relevant processes is also important.

\begin{figure}[!ht]
\includegraphics[width=0.8\linewidth]{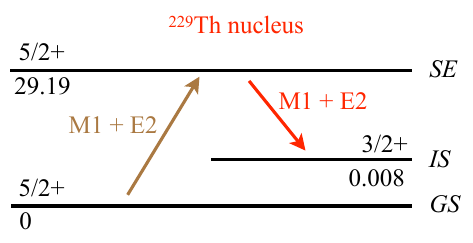}
\caption{\label{fig1-levles} 
Partial level scheme of $^{229}$Th. The nuclear isomeric ($IS$), second-excited ($SE$), and ground state ($GS$) levels are labeled by their spin, parity, and energy in keV.
}
\end{figure}

\begin{figure}[!ht]
\includegraphics[width=0.96\linewidth]{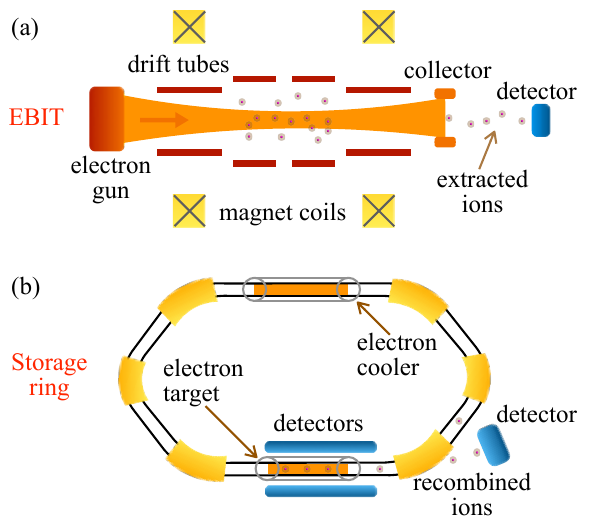}
\caption{\label{fig2-setup} 
(a) Sketch of an EBIT with ion extraction. (b) Sketch of a storage ring with an electron cooler and an electron target. Recombined ions can be recorded at the detector in the successive bending arc behind the electron target. X/$\gamma$-ray detectors are placed around the electron target for the detection of characteristic x/$\gamma$-ray photons of NEEC.
}
\end{figure}

Here we study theoretically the production of $^{229m}$Th via the excitation to the second-excited state at $29.19$ keV by NEEC: the nucleus is excited at first to the second-excited state by NEEC, and then decays to the isomeric state $^{229m}$Th with a branching ratio, as shown in Fig.~\ref{fig1-levles}. Our analyses are based on EBITs and storage rings, which could provide clean environments for the control of the isomer production and precise measurements of properties of the isomer, as well as NEEC observations. Sketches of an EBIT and a storage ring are presented in Fig.~\ref{fig2-setup}. In an EBIT \cite{DonetsRSI1990,MarrsPRL1988,LevinePS1988,GillaspyJPB2001,CurrellIEEE2005,BlessenohlRSI2018,MickeRSI2018, SchweigerRSI2019,BlaumPR2006,DillingARNPS2018,BlaumQST2021,KozlovRMP2018}, highly charged ions are produced by collisions with a tuneable and compressed electron beam. These highly charged ions are confined in the EBIT by the electron beam, the magnetic field generated by magnet coils, and the potentials in the trap. EBITs are powerful tools for the controllable production of highly charged ions for precise measurements \cite{BlaumPR2006,DillingARNPS2018,BlaumQST2021,KozlovRMP2018}. When tuning the electron energy to the resonant condition of a NEEC channel, NEEC may happen in the EBIT. After the occurrence of NEEC, the ions can be extracted from the EBIT for measurements. 

A different scenario involves storage rings \cite{GeisselPRL1992,XiaNIMPRA2022,LitvinovRPP2011,SchippersPRL2007,BrandauPS2013,GrieserEPJST2012, LitvinovNIMPRB2013,SteckPPNP2020,ZhouAAPPS2022}, where highly charged ions from an accelerator are stored in a ring. The energy spread of the ion beam can be cooled down by the electron beam of the electron cooler. An electron target, which is similar to the electron cooler, can be installed to the storage ring \cite{GrieserEPJST2012,LitvinovNIMPRB2013,SteckPPNP2020,ZhouAAPPS2022}. In such a scenario, one can trigger the NEEC process by tuning the electron energy of the electron target to the resonant condition of a NEEC channel. The recombined ions can be separated from the primary beam in the successive bending magnet for detection \cite{MaPS2015}, as in dielectronic recombination (DR) detections \cite{SchippersPRL2007,BrandauPS2013,GrieserEPJST2012,LitvinovNIMPRB2013,SteckPPNP2020,ZhouAAPPS2022}. X/$\gamma$-ray detectors can be placed around and a few meters after the electron target, which could help for identifying NEEC events.

Our calculations show that considering the parameters of such facilities, the production rate of $^{229m}$Th can reach $10^{-5}$ s$^{-1}$ per nucleus, which is about six orders of magnitude larger than the value experimentally demonstrated using $29$-keV synchrotron radiation (about $10^{-11}$ s$^{-1}$) for this indirect excitation \cite{MasudaNat2019}. We identify proper NEEC channels which can lead to the efficient production of $^{229m}$Th in an EBIT. By detecting $^{229m}$Th, the NEEC phenomenon can be confirmed. We also identify the proper NEEC channel in which a $100$ hours scan of the electron energy in a storage ring would be sufficient to observe a narrow peak in the number of recombined ions above the radiative recombination (RR) background for the observation of NEEC. Furthermore, we identify the characteristic signature of NEEC, i.e., the recombined ion and two photons (an x-ray photon from the atomic transition, followed by a $\gamma$ photon with a time delay of about $30$ ns from the radiative decay of the nuclear excited state). With this characteristic signature, the occurrence of NEEC could be followed and the NEEC event could be unambiguously identified.

\section{NEEC cross section}

The NEEC cross section of a given NEEC channel depends on the recombining electron energy $E$ mainly via a normalized Lorentz profile \cite{PalffyPRA2006,GunstPOP2015},
\begin{equation} \label{eq-sig}
  \sigma_{\rm{neec}} (E) = S_{\rm{neec}}  \frac{\Gamma/(2\pi)}{(E-E_r)^2 + \Gamma^2/4},
\end{equation}
where $S_{\rm{neec}}$ is the NEEC resonance strength, $E_r$ the resonance energy of the recombining electron, and $\Gamma$ the natural width of the resonant state. The NEEC rate for a given NEEC channel can be obtained by the convolution over the electron energy $E$ of the NEEC cross section and the electron flux distribution $\Phi_e (E)$,
\begin{equation}
  \lambda_{\rm{neec}} = \int dE \sigma_{\rm{neec}}(E)  \Phi_e(E).
\end{equation}

The NEEC cross sections and resonance strengths are calculated following the method developed in Ref.~\cite{PalffyPRA2006} and later extended and used for a number of NEEC studies under various scenarios \cite{PalffyPRA2007,PalffyPRL2007,GunstPRL2014,GunstPOP2015,Wu2019,PalffyPLB2008,WuPRL2018,WuPRL2022}. The multiconfigurational-Dirac-Fock method implemented in GRASP2018 \cite{FischerCPC2019} is employed to obtain the bound electronic wave functions. We obtain numerically the continuum electron wave functions from the solution of the Dirac equation with an effective charge. We note that so far the nuclear reduced transition probabilities for the first two excited states of $^{229}$Th have not yet been determined experimentally. In the present work, we use the necessary nuclear reduced transition probabilities from theoretical calculations in Refs.~\cite{MinkovPRL2017,MinkovPRL2019,MinkovPRC2021,KirschbaumPRC2022}. More precisely, we adopt the reduced transition probabilities $B({\rm{M1}}) = 0.0043$ W.u. (Weisskopf unit) and $B({\rm{E2}}) = 39.49$ W.u. for the transition from the nuclear second-excited state $|SE \rangle$ to the ground state $|GS \rangle$. Furthermore, the width $\Gamma$ in Eq.~(\ref{eq-sig}) for a given NEEC channel is the sum of the nuclear state width and the electronic width. For NEEC into the electronic ground state, the width $\Gamma$ is the nuclear state width, e.g., about $18$ neV when IC channels are not yet opened \cite{MasudaNat2019,SikorskyPRL2020,KirschbaumPRC2022}. However, for NEEC into an excited electronic configuration, the width $\Gamma$ can be determined by the electronic width. According to our GRASP calculations on the electron transitions, the width $\Gamma$ lies between the level of the nuclear state width and the level of $10$ eV, depending on the electronic configuration.

\begin{figure}[!ht]
\includegraphics[width=\linewidth]{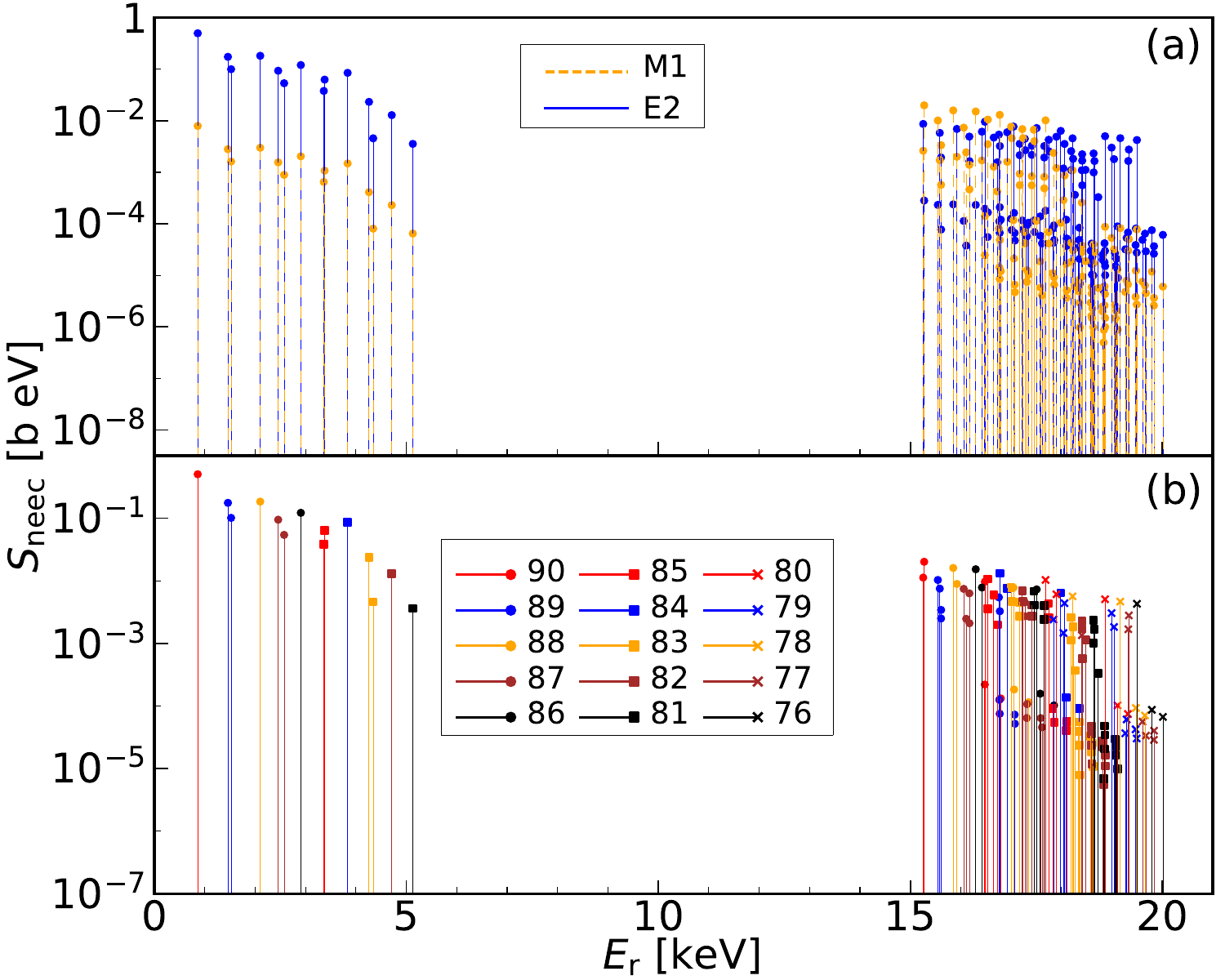}
\caption{\label{fig3-neecer} 
NEEC resonance strength $S_{\rm{neec}}$ for the transition from $|GS \rangle$ to $|SE \rangle$ of $^{229}$Th: (a) Contributions of ${\rm{M1}}$ transition and ${\rm{E2}}$ transition; (b) Total NEEC resonance strength. The initial ion is assumed to be in its electronic ground state with charge $q$ from $76$ to $90$. Results for allowed NEEC channels into $L$ shell and $M$ shell are presented.
}
\end{figure}

Figure~\ref{fig3-neecer} shows the NEEC resonance strength $S_{\rm{neec}}$ for the transition $|GS \rangle \rightarrow |SE \rangle$. We assume the initial ion is in its electronic ground state with charge $q$. For the initial ion we consider charge states $q$ from $76$ to $90$. Results of possible channels of NEEC into $L$ shell and $M$ shell are presented. Under the electronic ground state assumption of the initial ion, channels of NEEC into $1s_{1/2}$, $2s_{1/2}$, and $2p_{1/2}$ orbitals are forbidden for the transition $|GS\rangle \rightarrow |SE \rangle$, as binding energies of these electronic states exceed the nuclear transition energy. As shown in Fig.~\ref{fig3-neecer}, $S_{\rm{neec}}$ reaches the largest value of $0.51$ b eV for the initial ion charge state $q=90$.

Figure~\ref{fig3-neecer} also shows that the resonance electron energy locates in two regions: one in the range around $1$ keV to $5$ keV, another one in the range around $15$ keV to $20$ keV. The group of NEEC channels with low resonance electron energies is related to NEEC into the $2p_{3/2}$ orbital. And the group of NEEC channels with high resonance electron energies is related to NEEC into $M$ shell. As shown in Fig.~\ref{fig3-neecer}(a), for all considered NEEC channels of $2p_{3/2}$, the NEEC contribution from the ${\rm{E2}}$ transition dominates the one from the ${\rm{M1}}$ transition, although the $M1$ transition is much stronger than the ${\rm{E2}}$ transition for the radiative decay $|SE \rangle \rightarrow |GS \rangle$. This is because for ${\rm{E2}}$ transitions the NEEC channels into $p$ orbitals are particularly efficient. In contrast, the NEEC contribution from the ${\rm{M1}}$ transition dominates the one from the ${\rm{E2}}$ transition for the NEEC channels of $3s_{1/2}$.

After the occurrence of NEEC, the nucleus decays from the second-excited state $|SE \rangle$ to the isomeric state $|IS \rangle$ with a branching ratio. The branching ratio depends on the electronic configuration, as IC plays an important role. As in general the atomic transition rate is much higher than the IC rate (by orders of magnitude) for the case under consideration, after the occurrence of NEEC and before the occurrence of IC and radiative decay of the nucleus, the ion has transited to its electronic ground state. Thus, for charge states $q$ of the initial ion before NEEC higher than $84$, the contribution from IC channels for the decay of $|SE \rangle$ is negligible and only radiative decay plays a role (for charge states after NEEC higher or equal to $84$, the IC channels of $1s_{1/2}$, $2s_{1/2}$, and $2p_{1/2}$ orbitals are forbidden for the decay of $|SE \rangle$ when the ion is in its electronic ground state, as the electronic binding energies exceed the nuclear transition energy). The experiment in Ref.~\cite{SikorskyPRL2020} reported the radiative branching ratio of the in-band transition $|SE \rangle \rightarrow |IS \rangle$ as $\sim 90\%$. For charge states that allow the IC channels to contribute, the situation becomes complicated. In this case, due to the precision of the available experimental results and the inconsistencies between different experiments as well as theoretical predictions \cite{MasudaNat2019,SikorskyPRL2020,KirschbaumPRC2022}, there is a large uncertainty on the branching ratio of the transition $|SE \rangle \rightarrow |IS \rangle$. The branching ratio of the in-band transition is expected to be from around $60\%$ to $90\%$ when IC channels are involved. We expect that the uncertainty of the branching ratio does not change the order of magnitude for the production of $^{229m}$Th.

\section{NEEC in EBITs}

We are now turning to the analysis based on EBIT parameters. Assuming an electron beam with a current of around $100$ mA focused in a radius of $50$ $\mu$m with an energy spread of $10$ eV \cite{WangFP2023,MarrsPRL1988,LevinePS1988,BlessenohlRSI2018,MickeRSI2018, SchweigerRSI2019,BlaumPR2006,DillingARNPS2018,BlaumQST2021}, the electron flux in an EBIT can reach the level $10^{21}$ cm$^{-2}$ s$^{-1}$ eV$^{-1}$. Assuming a charge state of $90$ and the electron energy tuned to the resonance energy of the NEEC channel of $2p_{3/2}$, the NEEC rate is on the order of $10^{-4}$ s$^{-1}$ per ion. However, in order to product Th$^{90+}$ ions in an EBIT efficiently, the electron energy should be about $2.5$ times of the ionization energy of $K$ shell electrons of Th, which is about $300$ keV. As shown in Fig.~\ref{fig3-neecer}, the resonance energy of the NEEC channel of $2p_{3/2}$ at charge state $90$ is about $1$ keV. There is a large gap between these two energies. A two-electron-beam setup suggested in Ref.~\cite{WangFP2023} may be applicable: one high-energy electron beam for the production of Th$^{90+}$ ions, and another electron beam tuned to the resonance energy for NEEC. However, as the electron energy for the production of Th$^{90+}$ ions exceeds the excitation energy of the nuclear state $|SE \rangle$ as well as a number of nuclear excitation energies, inelastic scattering may also contribute to nuclear excitations. In addition, nuclear hyperfine mixing related to the isomeric state $^{229m}$Th in few-electron $^{229}$Th \cite{ShabaevPRL2022} may also play an important role in the production and decay of $^{229m}$Th. Thus, for controllable production of $^{229m}$Th as well as NEEC observation in EBITs, charge states closed to bare ions would not be a good choice.

As the ionization energies of $3s$ and $3p$ electrons of Th are around $10$ keV and the ionization energies of $K$ and $L$ shell electrons are higher than $20$ keV, an electron energy around $20$ keV is good for efficient production of Th ions at charge states between $76$ and $80$ \cite{SchneiderPRA1991}. As shown in Fig.~\ref{fig3-neecer}, the resonance energy of NEEC for charge states between $76$ and $80$ is also around $20$ keV. Thus, charge states between $76$ and $80$ should be good choices for NEEC in EBITs. We consider two NEEC channels: capturing into the $3s$ orbital of initial charge state $q=80$ with the resonance electron energy $17.46$ keV and $S_{\rm{neec}} = 1.0 \times 10^{-2}$ b eV, and capturing into the $3p_{1/2}$ orbital of initial charge state $q=78$ with the resonance electron energy $18.22$ keV and $S_{\rm{neec}} = 5.6 \times 10^{-3}$ b eV. When tuning the electron energy to the resonance energy of one of these two NEEC channels, the NEEC rate reaches $10^{-5}$ s$^{-1}$ per ion and $0.56 \times 10^{-5}$ s$^{-1}$ per ion, respectively. Assuming the total number of Th ions in the EBIT is $10^5$ and the electron energy of the beam applied for the production of Th ions is $24$ keV, the population of Th$^{80+}$ ions can reach $10\%$ and that of Th$^{78+}$ ions can reach $20\%$ \cite{SchneiderPRA1991}. Under this condition, the $^{229m}$Th production rate via NEEC reaches the order of $10^{-1}$ isomer per second. We note that, as the electron energies are lower than the excitation energy of the nuclear second excited state, NEEC should be the only opened known channel for the excitation to $|SE \rangle$ from the state $|GS \rangle$. Another possible process which may contribute to the production of $^{229m}$Th in such a scenario is inelastic electron scattering leading to the direct excitation from $|GS \rangle$ to $|IS \rangle$. According to Ref.~\cite{ZhangPRC2022}, the cross section of the inelastic electron scattering for the excitation $|GS \rangle \rightarrow |IS \rangle$ at the electron energy of $\sim 20$ keV is a few times of $10^{-30}$ cm$^2$. With the electron flux in the EBIT, the inelastic electron scattering rate is on the order of $10^{-7}$ s$^{-1}$ per ion. We note that the inelastic electron scattering is a non-resonant process that can happen for all charge states. Taking into account the charge state distribution for the entire ion cloud in the EBIT, the total NEEC event rate can still be larger than the total event rate of the inelastic electron scattering. Thus NEEC can be the dominating process for the production of $^{229m}$Th under such conditions. After the occurrence of NEEC, ions can be extracted from the EBIT for measurements by changing the potentials in the trap \cite{SchneiderPRA1991,SchneiderPRA1990,PenetrantePRA1991}. Assuming we extract all ions from the EBIT in every $1000$ s, about $0.1\%$ of the extracted ions should be in the nuclear isomeric state, and one can detect $^{229m}$Th as a signature of NEEC.

\section{NEEC in storage rings}

Now we turn to the scenario of storage rings, a clean environment which also has the potential of controllable production of $^{229m}$Th and clear observation of NEEC. In order to detect recombined ions in the downstream of the ion beam as NEEC signals as in dielectronic recombination, the charge state of the ion after the occurrence of NEEC should last sufficiently long. In this regard, IC channels for the decay of $|SE \rangle$ should be forbidden after the occurrence of NEEC, otherwise the ion charge state shifts back quickly to the one before the electron recombination. Thus, we focus on the cases that the contribution from IC channels for the decay of $|SE \rangle$ is negligible, i.e., initial charge state $q >84$. This is similar to the NEECX process studied in Ref.~\cite{PalffyPLB2008}. Furthermore, as mentioned above, the NEEC cross section connects to the resonance strength $S_{\rm{neec}}$ via a normalized Lorentz profile with the width $\Gamma$, and depending on the electronic configuration, the width $\Gamma$ can be from about $18$ neV to the level of $10$ eV. As the energy spread of ions and electrons can be very small (can be as low as mili-eV for electrons) \cite{XiaNIMPRA2022,LitvinovRPP2011,LitvinovNIMPRB2013,SteckPPNP2020}, a small width could be useful to overcome the background from the non-resonant radiative recombination process.

The most interesting case is the channel of NEEC into $2p_{3/2}$ for initial ion charge state $q=86$. In this channel, the NEEC resonance strength is $S_{\rm{neec}} = 0.12$ b eV, and the resonance electron energy is $2.907$ keV (energy in the ion rest frame). According to our GRASP calculation on electron transitions, the width $\Gamma$ for this NEEC channel is $\sim 0.1$ meV. After the occurrence of NEEC, the recombined $2p_{3/2}$ electron transits down quickly to the $2p_{1/2}$ orbital whose IC channel is forbidden for the decay of $|SE \rangle$. The main background of recombined ions should be from the RR process. Calculations by the FAC code \cite{GuCJP2008,facIaea2024} as well as following Refs.~\cite{PenetrantePRA1991,KimPRA1983} show that the RR cross section for the charge state $q=86$ at electron energy $2.9$ keV is $\approx 10^4$ b. Assuming the current of the electron beam is $100$ mA, the radius of the electron beam $1$ cm, and the energy spread on the level of mili-eV \cite{XiaNIMPRA2022,LitvinovRPP2011,LitvinovNIMPRB2013,SteckPPNP2020,MaPS2015}, the electron flux is on the order of $10^{20}$ cm$^{-2}$ s$^{-1}$ eV$^{-1}$ with the integrated electron flux $\sim 10^{17}$ cm$^{-2}$ s$^{-1}$. When tuning the electron energy of the electron target to the NEEC resonance energy, the NEEC rate is $\sim 10^{-5}$ s$^{-1}$ per ion, while the RR rate is $\sim 10^{-3}$ s$^{-1}$ per ion. We note that the interaction length (length of the electron target) is on the order of $1$ meter, and the length of the ring is on the order of $100$ meters. Assuming the number of ions in the ring is $10^8$, and counting the mentioned factor, a $100$ hours scan of the electron energy with the range of $10$ eV in the precision of $1$ meV would be sufficient to observe a narrow peak in the number of recombined ions above the RR background with the event number of a few hundreds. After the scan of the electron energy, the resonance energy of the NEEC channel can be determined with the precision of milli-eV. One can tune the electron energy precisely to the resonance energy for efficient and controllable production of $^{229m}$Th.

We note that as in the case of DR experiments \cite{SchippersPRL2007,BrandauPS2013,GrieserEPJST2012,LitvinovNIMPRB2013,SteckPPNP2020}, one can also use only the electron cooler for both the ion cooling and the electron source for NEEC. In this case, one has to tune the electron beam of the electron cooler to the same velocity of ions to cool the ion beam, and detune the electron energy of the electron cooler to the NEEC resonance energy for the occurrence of NEEC. After some time of interaction at the electron energy tuned to the NEEC resonance energy, the ion beam needs to be re-cooled in order to maintain the narrow energy spread by tuning the electron beam to the same velocity of ions. Additional efforts are required for the shifting between the two electron energies. In order to avoid such shifting of the electron energy, the scenario analysed in the present work involved an electron cooler and an electron target. In such a scenario, the narrow energy spread of the ion beam can be maintained by the electron cooler, and the NEEC resonance condition can be matched in the electron target.

We are now discussing another interesting feature of the NEEC channel of $2p_{3/2}$ for initial ion charge state $q=86$. After the occurrence of NEEC, the recombined $2p_{3/2}$ electron transits down quickly to the $2p_{1/2}$ orbital emitting an x-ray photon with energy $3.7$ keV, and followed by a $\gamma$ photon with energy $29.1$ keV with a time delay of about $30$ ns from the radiative decay of $|SE \rangle$ \cite{SikorskyPRL2020} (photon energies are energies in the ion rest frame). The two photons and recombined ion could be the characteristic signature of NEEC. We note that, as the relative electron energy is $2.907$ keV, and the total RR probability per cycle is about $10^{-3}$, the photon signals above the energy of $3$ keV should be clean. The x-ray photon together with the recombined ion should identify the recombination of an electron into the given atomic orbital, and the delay $\gamma$ photon should identify the nuclear excitation. Thus, the characteristic signature of NEEC offers the chance to follow the occurrence of NEEC, leading to the unambiguous identification of an NEEC event. For comparison, the most frequent signal suggested for NEEC observations is the signal from the nuclear excitation only, which could be induced by other completing nuclear excitation processes. With similar characteristic signature, other NEEC channels with initial ion charge state $q >84$ and larger width could also be used for NEEC observation. For example, for the NEEC channel of $2p_{3/2}$ for initial ion charge state $q=88$, the resonance electron energy is $2.101$ keV with width $\sim 5 \times 10^{-2}$ eV and $S_{\rm{neec}} = 0.18$ b eV. The characteristic photons in this case should be a $4.0$ keV photon followed by a $29.1$ keV photon.

\section{Conclusions}

We have studied in the present work the production of $^{229m}$Th by the excitation of $^{229}$Th to the above-lying excited state at $29.19$ keV energy via the process of NEEC. We have shown that for the parameters of EBITs and storage rings, the production rate of $^{229m}$Th can reach $10^{-5}$ s$^{-1}$ per nucleus. We have identified proper NEEC channels which offer the chance to confirm the NEEC phenomenon in EBITs and storage rings. Furthermore, we have also identified the recombined ion and a two-photon coincidence as the characteristic signature of NEEC: an x-ray photon from the atomic transition together with the recombined ion identifies the recombination of an electron into the given atomic orbital, and a following $\gamma$ photon with a time delay of about $30$ ns from the radiative decay of the nuclear excited state identifies the nuclear excitation.

The efficient and controllable production of $^{229m}$Th proposed in the present work, offers the chance for precise measurements of the properties of this unique isomer and isotope. Furthermore, as the NEEC process is very selective on the transition type depending on the NEEC channels, the NEEC observation would provide precise transition strengths for ${\rm{E2}}$ and ${\rm{M1}}$ transitions. This could offer a solution for the inconsistencies between different experiments as well as theoretical predictions \cite{MasudaNat2019,SikorskyPRL2020,KirschbaumPRC2022} in the transition strengths related to the first two excited states of $^{229}$Th. Furthermore, our analysis opens new opportunities of controllable production of isomers in EBITs and storage rings. The characteristic signature of NEEC identified in the present work, which could allow us to follow the occurrence of NEEC to unambiguously identify the NEEC event, also opens an excellent path for the confirmation of the long-sought NEEC phenomenon.

\begin{acknowledgements}
The authors thank Song Guo for fruitful discussions. This work was supported by the Fundamental Research Funds for the Central Universities (Grants No. 010-63233008 and No. 010-63243088). AP gratefully acknowledges support from the Deutsche Forschungsgemeinschaft (DFG, German Science Foundation) in the framework of the Heisenberg Program (PA 2508/3-1) and from the ThoriumNuclearClock project that has received funding from the European Research Council (ERC) under the European Union's Horizon 2020 research and innovation programme (Grant Agreement No. 856415).
\end{acknowledgements}



\bibliographystyle{apsrev-no-url-issn}
\bibliography{refs229mth}{}

\begin{thebibliography}{105}
\expandafter\ifx\csname natexlab\endcsname\relax\def\natexlab#1{#1}\fi
\expandafter\ifx\csname bibnamefont\endcsname\relax
  \def\bibnamefont#1{#1}\fi
\expandafter\ifx\csname bibfnamefont\endcsname\relax
  \def\bibfnamefont#1{#1}\fi
\expandafter\ifx\csname citenamefont\endcsname\relax
  \def\citenamefont#1{#1}\fi
\expandafter\ifx\csname url\endcsname\relax
  \def\url#1{\texttt{#1}}\fi
\expandafter\ifx\csname urlprefix\endcsname\relax\def\urlprefix{URL }\fi
\providecommand{\bibinfo}[2]{#2}
\providecommand{\eprint}[2][]{\url{#2}}

\bibitem[{\citenamefont{Walker and Dracoulis}(1999)}]{Walker1999}
\bibinfo{author}{\bibfnamefont{P.}~\bibnamefont{Walker}} \bibnamefont{and}
  \bibinfo{author}{\bibfnamefont{G.}~\bibnamefont{Dracoulis}},
  \bibinfo{journal}{Nature} \textbf{\bibinfo{volume}{399}}, \bibinfo{pages}{35}
  (\bibinfo{year}{1999}).

\bibitem[{\citenamefont{Walker and Podoly{\'ak}}(2020)}]{Walker2020}
\bibinfo{author}{\bibfnamefont{P.}~\bibnamefont{Walker}} \bibnamefont{and}
  \bibinfo{author}{\bibfnamefont{Z.}~\bibnamefont{Podoly{\'ak}}},
  \bibinfo{journal}{Physica Scripta} \textbf{\bibinfo{volume}{95}},
  \bibinfo{pages}{044004} (\bibinfo{year}{2020}).

\bibitem[{\citenamefont{Peik and Tamm}(2003)}]{PeikEPL2003}
\bibinfo{author}{\bibfnamefont{E.}~\bibnamefont{Peik}} \bibnamefont{and}
  \bibinfo{author}{\bibfnamefont{C.}~\bibnamefont{Tamm}},
  \bibinfo{journal}{Europhysics Letters} \textbf{\bibinfo{volume}{61}},
  \bibinfo{pages}{181} (\bibinfo{year}{2003}).

\bibitem[{\citenamefont{von~der Wense and Seiferle}(2020)}]{WenseEPJA2020}
\bibinfo{author}{\bibfnamefont{L.}~\bibnamefont{von~der Wense}}
  \bibnamefont{and} \bibinfo{author}{\bibfnamefont{B.}~\bibnamefont{Seiferle}},
  \bibinfo{journal}{Eur. Phys. J. A} \textbf{\bibinfo{volume}{56}},
  \bibinfo{pages}{277} (\bibinfo{year}{2020}).

\bibitem[{\citenamefont{Beeks et~al.}(2021)\citenamefont{Beeks, Sikorsky,
  Schumm, Thielking, Okhapkin, and Peik}}]{BeeksNRP2021}
\bibinfo{author}{\bibfnamefont{K.}~\bibnamefont{Beeks}},
  \bibinfo{author}{\bibfnamefont{T.}~\bibnamefont{Sikorsky}},
  \bibinfo{author}{\bibfnamefont{T.}~\bibnamefont{Schumm}},
  \bibinfo{author}{\bibfnamefont{J.}~\bibnamefont{Thielking}},
  \bibinfo{author}{\bibfnamefont{M.~V.} \bibnamefont{Okhapkin}},
  \bibnamefont{and} \bibinfo{author}{\bibfnamefont{E.}~\bibnamefont{Peik}},
  \bibinfo{journal}{Nature Reviews Physics} \textbf{\bibinfo{volume}{3}},
  \bibinfo{pages}{238} (\bibinfo{year}{2021}).

\bibitem[{\citenamefont{Campbell et~al.}(2012)\citenamefont{Campbell, Radnaev,
  Kuzmich, Dzuba, Flambaum, and Derevianko}}]{CampbellPRL2012}
\bibinfo{author}{\bibfnamefont{C.~J.} \bibnamefont{Campbell}},
  \bibinfo{author}{\bibfnamefont{A.~G.} \bibnamefont{Radnaev}},
  \bibinfo{author}{\bibfnamefont{A.}~\bibnamefont{Kuzmich}},
  \bibinfo{author}{\bibfnamefont{V.~A.} \bibnamefont{Dzuba}},
  \bibinfo{author}{\bibfnamefont{V.~V.} \bibnamefont{Flambaum}},
  \bibnamefont{and}
  \bibinfo{author}{\bibfnamefont{A.}~\bibnamefont{Derevianko}},
  \bibinfo{journal}{Phys. Rev. Lett.} \textbf{\bibinfo{volume}{108}},
  \bibinfo{pages}{120802} (\bibinfo{year}{2012}).

\bibitem[{\citenamefont{Kazakov et~al.}(2012)\citenamefont{Kazakov, Litvinov,
  Romanenko, Yatsenko, Romanenko, Schreitl, Winkler, and
  Schumm}}]{KazakovNJP2012}
\bibinfo{author}{\bibfnamefont{G.~A.} \bibnamefont{Kazakov}},
  \bibinfo{author}{\bibfnamefont{A.~N.} \bibnamefont{Litvinov}},
  \bibinfo{author}{\bibfnamefont{V.~I.} \bibnamefont{Romanenko}},
  \bibinfo{author}{\bibfnamefont{L.~P.} \bibnamefont{Yatsenko}},
  \bibinfo{author}{\bibfnamefont{A.~V.} \bibnamefont{Romanenko}},
  \bibinfo{author}{\bibfnamefont{M.}~\bibnamefont{Schreitl}},
  \bibinfo{author}{\bibfnamefont{G.}~\bibnamefont{Winkler}}, \bibnamefont{and}
  \bibinfo{author}{\bibfnamefont{T.}~\bibnamefont{Schumm}},
  \bibinfo{journal}{New J. Phys.} \textbf{\bibinfo{volume}{14}},
  \bibinfo{pages}{083019} (\bibinfo{year}{2012}).

\bibitem[{\citenamefont{Peik et~al.}(2021)\citenamefont{Peik, Schumm,
  Safronova, P\'alffy, Weitenberg, and Thirolf}}]{PeikQST2021}
\bibinfo{author}{\bibfnamefont{E.}~\bibnamefont{Peik}},
  \bibinfo{author}{\bibfnamefont{T.}~\bibnamefont{Schumm}},
  \bibinfo{author}{\bibfnamefont{M.~S.} \bibnamefont{Safronova}},
  \bibinfo{author}{\bibfnamefont{A.}~\bibnamefont{P\'alffy}},
  \bibinfo{author}{\bibfnamefont{J.}~\bibnamefont{Weitenberg}},
  \bibnamefont{and} \bibinfo{author}{\bibfnamefont{P.~G.}
  \bibnamefont{Thirolf}}, \bibinfo{journal}{Quantum Sci. Technol.}
  \textbf{\bibinfo{volume}{6}}, \bibinfo{pages}{034002} (\bibinfo{year}{2021}).

\bibitem[{\citenamefont{Flambaum}(2006)}]{FlambaumPRL2006}
\bibinfo{author}{\bibfnamefont{V.~V.} \bibnamefont{Flambaum}},
  \bibinfo{journal}{Phys. Rev. Lett.} \textbf{\bibinfo{volume}{97}},
  \bibinfo{pages}{092502} (\bibinfo{year}{2006}).

\bibitem[{\citenamefont{Flambaum}(2016)}]{FlambaumPRL2016}
\bibinfo{author}{\bibfnamefont{V.~V.} \bibnamefont{Flambaum}},
  \bibinfo{journal}{Phys. Rev. Lett.} \textbf{\bibinfo{volume}{117}},
  \bibinfo{pages}{072501} (\bibinfo{year}{2016}).

\bibitem[{\citenamefont{Fadeev et~al.}(2020)\citenamefont{Fadeev, Berengut, and
  Flambaum}}]{FadeevPRA2020}
\bibinfo{author}{\bibfnamefont{P.}~\bibnamefont{Fadeev}},
  \bibinfo{author}{\bibfnamefont{J.~C.} \bibnamefont{Berengut}},
  \bibnamefont{and} \bibinfo{author}{\bibfnamefont{V.~V.}
  \bibnamefont{Flambaum}}, \bibinfo{journal}{Phys. Rev. A}
  \textbf{\bibinfo{volume}{102}}, \bibinfo{pages}{052833}
  (\bibinfo{year}{2020}).

\bibitem[{\citenamefont{Tkalya}(2011)}]{TkalyaPRL2011}
\bibinfo{author}{\bibfnamefont{E.~V.} \bibnamefont{Tkalya}},
  \bibinfo{journal}{Phys. Rev. Lett.} \textbf{\bibinfo{volume}{106}},
  \bibinfo{pages}{162501} (\bibinfo{year}{2011}).

\bibitem[{\citenamefont{Thirolf et~al.}(2019)\citenamefont{Thirolf, Seiferle,
  and von~der Wense}}]{ThirolfAP2019}
\bibinfo{author}{\bibfnamefont{P.~G.} \bibnamefont{Thirolf}},
  \bibinfo{author}{\bibfnamefont{B.}~\bibnamefont{Seiferle}}, \bibnamefont{and}
  \bibinfo{author}{\bibfnamefont{L.}~\bibnamefont{von~der Wense}},
  \bibinfo{journal}{Ann. Phys. (Berlin)} \textbf{\bibinfo{volume}{531}},
  \bibinfo{pages}{1800381} (\bibinfo{year}{2019}).

\bibitem[{\citenamefont{von~der Wense et~al.}(2016)\citenamefont{von~der Wense,
  Seiferle, Laatiaoui, Neumayr, Maier, Wirth, Mokry, Runke, Eberhardt,
  D\"ullmann et~al.}}]{WenseNat2016}
\bibinfo{author}{\bibfnamefont{L.}~\bibnamefont{von~der Wense}},
  \bibinfo{author}{\bibfnamefont{B.}~\bibnamefont{Seiferle}},
  \bibinfo{author}{\bibfnamefont{M.}~\bibnamefont{Laatiaoui}},
  \bibinfo{author}{\bibfnamefont{J.~B.} \bibnamefont{Neumayr}},
  \bibinfo{author}{\bibfnamefont{H.-J.} \bibnamefont{Maier}},
  \bibinfo{author}{\bibfnamefont{H.-F.} \bibnamefont{Wirth}},
  \bibinfo{author}{\bibfnamefont{C.}~\bibnamefont{Mokry}},
  \bibinfo{author}{\bibfnamefont{J.}~\bibnamefont{Runke}},
  \bibinfo{author}{\bibfnamefont{K.}~\bibnamefont{Eberhardt}},
  \bibinfo{author}{\bibfnamefont{C.~E.} \bibnamefont{D\"ullmann}},
  \bibnamefont{et~al.}, \bibinfo{journal}{Nature (London)}
  \textbf{\bibinfo{volume}{533}}, \bibinfo{pages}{47} (\bibinfo{year}{2016}).

\bibitem[{\citenamefont{Thielking et~al.}(2018)\citenamefont{Thielking,
  Okhapkin, G\l{}owacki, Meier, von~der Wense, Seiferle, D\"ullmann, Thirolf,
  and Peik}}]{ThielkingNat2018}
\bibinfo{author}{\bibfnamefont{J.}~\bibnamefont{Thielking}},
  \bibinfo{author}{\bibfnamefont{M.~V.} \bibnamefont{Okhapkin}},
  \bibinfo{author}{\bibfnamefont{P.}~\bibnamefont{G\l{}owacki}},
  \bibinfo{author}{\bibfnamefont{D.~M.} \bibnamefont{Meier}},
  \bibinfo{author}{\bibfnamefont{L.}~\bibnamefont{von~der Wense}},
  \bibinfo{author}{\bibfnamefont{B.}~\bibnamefont{Seiferle}},
  \bibinfo{author}{\bibfnamefont{C.~E.} \bibnamefont{D\"ullmann}},
  \bibinfo{author}{\bibfnamefont{P.~G.} \bibnamefont{Thirolf}},
  \bibnamefont{and} \bibinfo{author}{\bibfnamefont{E.}~\bibnamefont{Peik}},
  \bibinfo{journal}{Nature (London)} \textbf{\bibinfo{volume}{556}},
  \bibinfo{pages}{321} (\bibinfo{year}{2018}).

\bibitem[{\citenamefont{Seiferle et~al.}(2017)\citenamefont{Seiferle, von~der
  Wense, and Thirolf}}]{SeiferlePRL2017}
\bibinfo{author}{\bibfnamefont{B.}~\bibnamefont{Seiferle}},
  \bibinfo{author}{\bibfnamefont{L.}~\bibnamefont{von~der Wense}},
  \bibnamefont{and} \bibinfo{author}{\bibfnamefont{P.~G.}
  \bibnamefont{Thirolf}}, \bibinfo{journal}{Phys. Rev. Lett.}
  \textbf{\bibinfo{volume}{118}}, \bibinfo{pages}{042501}
  (\bibinfo{year}{2017}).

\bibitem[{\citenamefont{Seiferle et~al.}(2019)\citenamefont{Seiferle, von~der
  Wense, Bilous, Amersdorffer, Lemell, Libisch, Stellmer, Schumm, D\"ullmann,
  P\'alffy et~al.}}]{SeiferleNat2019}
\bibinfo{author}{\bibfnamefont{B.}~\bibnamefont{Seiferle}},
  \bibinfo{author}{\bibfnamefont{L.}~\bibnamefont{von~der Wense}},
  \bibinfo{author}{\bibfnamefont{P.~V.} \bibnamefont{Bilous}},
  \bibinfo{author}{\bibfnamefont{I.}~\bibnamefont{Amersdorffer}},
  \bibinfo{author}{\bibfnamefont{C.}~\bibnamefont{Lemell}},
  \bibinfo{author}{\bibfnamefont{F.}~\bibnamefont{Libisch}},
  \bibinfo{author}{\bibfnamefont{S.}~\bibnamefont{Stellmer}},
  \bibinfo{author}{\bibfnamefont{T.}~\bibnamefont{Schumm}},
  \bibinfo{author}{\bibfnamefont{C.~E.} \bibnamefont{D\"ullmann}},
  \bibinfo{author}{\bibfnamefont{A.}~\bibnamefont{P\'alffy}},
  \bibnamefont{et~al.}, \bibinfo{journal}{Nature (London)}
  \textbf{\bibinfo{volume}{573}}, \bibinfo{pages}{243} (\bibinfo{year}{2019}).

\bibitem[{\citenamefont{Yamaguchi et~al.}(2019)\citenamefont{Yamaguchi,
  Muramatsu, Hayashi, Yuasa, Nakamura, Takimoto, Haba, Konashi, Watanabe,
  Kikunaga et~al.}}]{YamaguchiPRL2019}
\bibinfo{author}{\bibfnamefont{A.}~\bibnamefont{Yamaguchi}},
  \bibinfo{author}{\bibfnamefont{H.}~\bibnamefont{Muramatsu}},
  \bibinfo{author}{\bibfnamefont{T.}~\bibnamefont{Hayashi}},
  \bibinfo{author}{\bibfnamefont{N.}~\bibnamefont{Yuasa}},
  \bibinfo{author}{\bibfnamefont{K.}~\bibnamefont{Nakamura}},
  \bibinfo{author}{\bibfnamefont{M.}~\bibnamefont{Takimoto}},
  \bibinfo{author}{\bibfnamefont{H.}~\bibnamefont{Haba}},
  \bibinfo{author}{\bibfnamefont{K.}~\bibnamefont{Konashi}},
  \bibinfo{author}{\bibfnamefont{M.}~\bibnamefont{Watanabe}},
  \bibinfo{author}{\bibfnamefont{H.}~\bibnamefont{Kikunaga}},
  \bibnamefont{et~al.}, \bibinfo{journal}{Phys. Rev. Lett.}
  \textbf{\bibinfo{volume}{123}}, \bibinfo{pages}{222501}
  (\bibinfo{year}{2019}).

\bibitem[{\citenamefont{Sikorsky et~al.}(2020)\citenamefont{Sikorsky, Geist,
  Hengstler, Kempf, Gastaldo, Enss, Mokry, Runke, D\"ullmann, Wobrauschek
  et~al.}}]{SikorskyPRL2020}
\bibinfo{author}{\bibfnamefont{T.}~\bibnamefont{Sikorsky}},
  \bibinfo{author}{\bibfnamefont{J.}~\bibnamefont{Geist}},
  \bibinfo{author}{\bibfnamefont{D.}~\bibnamefont{Hengstler}},
  \bibinfo{author}{\bibfnamefont{S.}~\bibnamefont{Kempf}},
  \bibinfo{author}{\bibfnamefont{L.}~\bibnamefont{Gastaldo}},
  \bibinfo{author}{\bibfnamefont{C.}~\bibnamefont{Enss}},
  \bibinfo{author}{\bibfnamefont{C.}~\bibnamefont{Mokry}},
  \bibinfo{author}{\bibfnamefont{J.}~\bibnamefont{Runke}},
  \bibinfo{author}{\bibfnamefont{C.~E.} \bibnamefont{D\"ullmann}},
  \bibinfo{author}{\bibfnamefont{P.}~\bibnamefont{Wobrauschek}},
  \bibnamefont{et~al.}, \bibinfo{journal}{Phys. Rev. Lett.}
  \textbf{\bibinfo{volume}{125}}, \bibinfo{pages}{142503}
  (\bibinfo{year}{2020}).

\bibitem[{\citenamefont{Yamaguchi et~al.}(2024)\citenamefont{Yamaguchi,
  Shigekawa, Haba, Kikunaga, Shirasaki, Wada, and Katori}}]{YamaguchiNat2024}
\bibinfo{author}{\bibfnamefont{A.}~\bibnamefont{Yamaguchi}},
  \bibinfo{author}{\bibfnamefont{Y.}~\bibnamefont{Shigekawa}},
  \bibinfo{author}{\bibfnamefont{H.}~\bibnamefont{Haba}},
  \bibinfo{author}{\bibfnamefont{H.}~\bibnamefont{Kikunaga}},
  \bibinfo{author}{\bibfnamefont{K.}~\bibnamefont{Shirasaki}},
  \bibinfo{author}{\bibfnamefont{M.}~\bibnamefont{Wada}}, \bibnamefont{and}
  \bibinfo{author}{\bibfnamefont{H.}~\bibnamefont{Katori}},
  \bibinfo{journal}{Nature (London)} \textbf{\bibinfo{volume}{629}},
  \bibinfo{pages}{62} (\bibinfo{year}{2024}).

\bibitem[{\citenamefont{Kraemer et~al.}(2023)\citenamefont{Kraemer, Moens,
  Athanasakis-Kaklamanakis, Bara, Beeks, Chhetri, Chrysalidis, Claessens,
  Cocolios, Correia et~al.}}]{KraemerNat2023}
\bibinfo{author}{\bibfnamefont{S.}~\bibnamefont{Kraemer}},
  \bibinfo{author}{\bibfnamefont{J.}~\bibnamefont{Moens}},
  \bibinfo{author}{\bibfnamefont{M.}~\bibnamefont{Athanasakis-Kaklamanakis}},
  \bibinfo{author}{\bibfnamefont{S.}~\bibnamefont{Bara}},
  \bibinfo{author}{\bibfnamefont{K.}~\bibnamefont{Beeks}},
  \bibinfo{author}{\bibfnamefont{P.}~\bibnamefont{Chhetri}},
  \bibinfo{author}{\bibfnamefont{K.}~\bibnamefont{Chrysalidis}},
  \bibinfo{author}{\bibfnamefont{A.}~\bibnamefont{Claessens}},
  \bibinfo{author}{\bibfnamefont{T.~E.} \bibnamefont{Cocolios}},
  \bibinfo{author}{\bibfnamefont{J.~G.~M.} \bibnamefont{Correia}},
  \bibnamefont{et~al.}, \bibinfo{journal}{Nature (London)}
  \textbf{\bibinfo{volume}{617}}, \bibinfo{pages}{706} (\bibinfo{year}{2023}).

\bibitem[{\citenamefont{Masuda et~al.}(2019)\citenamefont{Masuda, Yoshimi,
  Fujieda, Fujimoto, Haba, Hara, Hiraki, Kaino, Kasamatsu, Kitao
  et~al.}}]{MasudaNat2019}
\bibinfo{author}{\bibfnamefont{T.}~\bibnamefont{Masuda}},
  \bibinfo{author}{\bibfnamefont{A.}~\bibnamefont{Yoshimi}},
  \bibinfo{author}{\bibfnamefont{A.}~\bibnamefont{Fujieda}},
  \bibinfo{author}{\bibfnamefont{H.}~\bibnamefont{Fujimoto}},
  \bibinfo{author}{\bibfnamefont{H.}~\bibnamefont{Haba}},
  \bibinfo{author}{\bibfnamefont{H.}~\bibnamefont{Hara}},
  \bibinfo{author}{\bibfnamefont{T.}~\bibnamefont{Hiraki}},
  \bibinfo{author}{\bibfnamefont{H.}~\bibnamefont{Kaino}},
  \bibinfo{author}{\bibfnamefont{Y.}~\bibnamefont{Kasamatsu}},
  \bibinfo{author}{\bibfnamefont{S.}~\bibnamefont{Kitao}},
  \bibnamefont{et~al.}, \bibinfo{journal}{Nature (London)}
  \textbf{\bibinfo{volume}{573}}, \bibinfo{pages}{238} (\bibinfo{year}{2019}).

\bibitem[{\citenamefont{Tiedau et~al.}(2024)\citenamefont{Tiedau, Okhapkin,
  Zhang, Thielking, Zitzer, Peik, Schaden, Pronebner, Morawetz, De~Col
  et~al.}}]{TiedauPRL2024}
\bibinfo{author}{\bibfnamefont{J.}~\bibnamefont{Tiedau}},
  \bibinfo{author}{\bibfnamefont{M.~V.} \bibnamefont{Okhapkin}},
  \bibinfo{author}{\bibfnamefont{K.}~\bibnamefont{Zhang}},
  \bibinfo{author}{\bibfnamefont{J.}~\bibnamefont{Thielking}},
  \bibinfo{author}{\bibfnamefont{G.}~\bibnamefont{Zitzer}},
  \bibinfo{author}{\bibfnamefont{E.}~\bibnamefont{Peik}},
  \bibinfo{author}{\bibfnamefont{F.}~\bibnamefont{Schaden}},
  \bibinfo{author}{\bibfnamefont{T.}~\bibnamefont{Pronebner}},
  \bibinfo{author}{\bibfnamefont{I.}~\bibnamefont{Morawetz}},
  \bibinfo{author}{\bibfnamefont{L.~T.} \bibnamefont{De~Col}},
  \bibnamefont{et~al.}, \bibinfo{journal}{Phys. Rev. Lett.}
  \textbf{\bibinfo{volume}{132}}, \bibinfo{pages}{182501}
  (\bibinfo{year}{2024}).

\bibitem[{\citenamefont{Elwell et~al.}(2024)\citenamefont{Elwell, Schneider,
  Jeet, Terhune, Morgan, Alexandrova, Derevianko, and Hudson}}]{ElwellPRL2024}
\bibinfo{author}{\bibfnamefont{R.}~\bibnamefont{Elwell}},
  \bibinfo{author}{\bibfnamefont{C.}~\bibnamefont{Schneider}},
  \bibinfo{author}{\bibfnamefont{J.}~\bibnamefont{Jeet}},
  \bibinfo{author}{\bibfnamefont{J.~E.~S.} \bibnamefont{Terhune}},
  \bibinfo{author}{\bibfnamefont{H.~W.~T.} \bibnamefont{Morgan}},
  \bibinfo{author}{\bibfnamefont{H.~B.~T.} \bibnamefont{Alexandrova},
  \bibfnamefont{A.~N.~and}},
  \bibinfo{author}{\bibfnamefont{A.}~\bibnamefont{Derevianko}},
  \bibnamefont{and} \bibinfo{author}{\bibfnamefont{E.~R.}
  \bibnamefont{Hudson}}, \bibinfo{journal}{arXiv: 2404.12311}
  (\bibinfo{year}{2024}).

\bibitem[{\citenamefont{von~der Wense et~al.}(2017)\citenamefont{von~der Wense,
  Seiferle, Stellmer, Weitenberg, Kazakov, P\'alffy, and
  Thirolf}}]{WensePRL2017}
\bibinfo{author}{\bibfnamefont{L.}~\bibnamefont{von~der Wense}},
  \bibinfo{author}{\bibfnamefont{B.}~\bibnamefont{Seiferle}},
  \bibinfo{author}{\bibfnamefont{S.}~\bibnamefont{Stellmer}},
  \bibinfo{author}{\bibfnamefont{J.}~\bibnamefont{Weitenberg}},
  \bibinfo{author}{\bibfnamefont{G.}~\bibnamefont{Kazakov}},
  \bibinfo{author}{\bibfnamefont{A.}~\bibnamefont{P\'alffy}}, \bibnamefont{and}
  \bibinfo{author}{\bibfnamefont{P.~G.} \bibnamefont{Thirolf}},
  \bibinfo{journal}{Phys. Rev. Lett.} \textbf{\bibinfo{volume}{119}},
  \bibinfo{pages}{132503} (\bibinfo{year}{2017}).

\bibitem[{\citenamefont{Kirschbaum et~al.}(2022)\citenamefont{Kirschbaum,
  Minkov, and P\'alffy}}]{KirschbaumPRC2022}
\bibinfo{author}{\bibfnamefont{T.}~\bibnamefont{Kirschbaum}},
  \bibinfo{author}{\bibfnamefont{N.}~\bibnamefont{Minkov}}, \bibnamefont{and}
  \bibinfo{author}{\bibfnamefont{A.}~\bibnamefont{P\'alffy}},
  \bibinfo{journal}{Phys. Rev. C} \textbf{\bibinfo{volume}{105}},
  \bibinfo{pages}{064313} (\bibinfo{year}{2022}).

\bibitem[{\citenamefont{Jin et~al.}(2023)\citenamefont{Jin, Bekker, Kirschbaum,
  Litvinov, P\'alffy, Sommerfeldt, Surzhykov, Thirolf, and
  Budker}}]{JinPRR2023}
\bibinfo{author}{\bibfnamefont{J.}~\bibnamefont{Jin}},
  \bibinfo{author}{\bibfnamefont{H.}~\bibnamefont{Bekker}},
  \bibinfo{author}{\bibfnamefont{T.}~\bibnamefont{Kirschbaum}},
  \bibinfo{author}{\bibfnamefont{Y.~A.} \bibnamefont{Litvinov}},
  \bibinfo{author}{\bibfnamefont{A.}~\bibnamefont{P\'alffy}},
  \bibinfo{author}{\bibfnamefont{J.}~\bibnamefont{Sommerfeldt}},
  \bibinfo{author}{\bibfnamefont{A.}~\bibnamefont{Surzhykov}},
  \bibinfo{author}{\bibfnamefont{P.~G.} \bibnamefont{Thirolf}},
  \bibnamefont{and} \bibinfo{author}{\bibfnamefont{D.}~\bibnamefont{Budker}},
  \bibinfo{journal}{Phys. Rev. Res.} \textbf{\bibinfo{volume}{5}},
  \bibinfo{pages}{023134} (\bibinfo{year}{2023}).

\bibitem[{\citenamefont{Tkalya}(2022)}]{TkalyaNPA2022}
\bibinfo{author}{\bibfnamefont{E.~V.} \bibnamefont{Tkalya}},
  \bibinfo{journal}{Nucl. Phys. A} \textbf{\bibinfo{volume}{1022}},
  \bibinfo{pages}{122428} (\bibinfo{year}{2022}).

\bibitem[{\citenamefont{Tkalya}(2020)}]{TkalyaPRL2020}
\bibinfo{author}{\bibfnamefont{E.~V.} \bibnamefont{Tkalya}},
  \bibinfo{journal}{Phys. Rev. Lett.} \textbf{\bibinfo{volume}{124}},
  \bibinfo{pages}{242501} (\bibinfo{year}{2020}).

\bibitem[{\citenamefont{Wang et~al.}(2021)\citenamefont{Wang, Zhou, Liu, and
  Wang}}]{WangPRL2021}
\bibinfo{author}{\bibfnamefont{W.}~\bibnamefont{Wang}},
  \bibinfo{author}{\bibfnamefont{J.}~\bibnamefont{Zhou}},
  \bibinfo{author}{\bibfnamefont{B.}~\bibnamefont{Liu}}, \bibnamefont{and}
  \bibinfo{author}{\bibfnamefont{X.}~\bibnamefont{Wang}},
  \bibinfo{journal}{Phys. Rev. Lett.} \textbf{\bibinfo{volume}{127}},
  \bibinfo{pages}{052501} (\bibinfo{year}{2021}).

\bibitem[{\citenamefont{Qi et~al.}(2023)\citenamefont{Qi, Zhang, and
  Wang}}]{QiPRL2023}
\bibinfo{author}{\bibfnamefont{J.}~\bibnamefont{Qi}},
  \bibinfo{author}{\bibfnamefont{H.}~\bibnamefont{Zhang}}, \bibnamefont{and}
  \bibinfo{author}{\bibfnamefont{X.}~\bibnamefont{Wang}},
  \bibinfo{journal}{Phys. Rev. Lett.} \textbf{\bibinfo{volume}{130}},
  \bibinfo{pages}{112501} (\bibinfo{year}{2023}).

\bibitem[{\citenamefont{Porsev and Flambaum}(2010)}]{PorsevPRA2010}
\bibinfo{author}{\bibfnamefont{S.~G.} \bibnamefont{Porsev}} \bibnamefont{and}
  \bibinfo{author}{\bibfnamefont{V.~V.} \bibnamefont{Flambaum}},
  \bibinfo{journal}{Phys. Rev. A} \textbf{\bibinfo{volume}{81}},
  \bibinfo{pages}{032504} (\bibinfo{year}{2010}).

\bibitem[{\citenamefont{Nickerson et~al.}(2020)\citenamefont{Nickerson, Pimon,
  Bilous, Gugler, Beeks, Sikorsky, Mohn, Schumm, and
  P\'alffy}}]{NickersonPRL2020}
\bibinfo{author}{\bibfnamefont{B.~S.} \bibnamefont{Nickerson}},
  \bibinfo{author}{\bibfnamefont{M.}~\bibnamefont{Pimon}},
  \bibinfo{author}{\bibfnamefont{P.~V.} \bibnamefont{Bilous}},
  \bibinfo{author}{\bibfnamefont{J.}~\bibnamefont{Gugler}},
  \bibinfo{author}{\bibfnamefont{K.}~\bibnamefont{Beeks}},
  \bibinfo{author}{\bibfnamefont{T.}~\bibnamefont{Sikorsky}},
  \bibinfo{author}{\bibfnamefont{P.}~\bibnamefont{Mohn}},
  \bibinfo{author}{\bibfnamefont{T.}~\bibnamefont{Schumm}}, \bibnamefont{and}
  \bibinfo{author}{\bibfnamefont{A.}~\bibnamefont{P\'alffy}},
  \bibinfo{journal}{Phys. Rev. Lett.} \textbf{\bibinfo{volume}{125}},
  \bibinfo{pages}{032501} (\bibinfo{year}{2020}).

\bibitem[{\citenamefont{Bilous et~al.}(2020)\citenamefont{Bilous, Bekker,
  Berengut, Seiferle, von~der Wense, Thirolf, Pfeifer, L\'opez-Urrutia, and
  P\'alffy}}]{BilousPRL2020}
\bibinfo{author}{\bibfnamefont{P.~V.} \bibnamefont{Bilous}},
  \bibinfo{author}{\bibfnamefont{H.}~\bibnamefont{Bekker}},
  \bibinfo{author}{\bibfnamefont{J.~C.} \bibnamefont{Berengut}},
  \bibinfo{author}{\bibfnamefont{B.}~\bibnamefont{Seiferle}},
  \bibinfo{author}{\bibfnamefont{L.}~\bibnamefont{von~der Wense}},
  \bibinfo{author}{\bibfnamefont{P.~G.} \bibnamefont{Thirolf}},
  \bibinfo{author}{\bibfnamefont{T.}~\bibnamefont{Pfeifer}},
  \bibinfo{author}{\bibfnamefont{J.~R.~C.} \bibnamefont{L\'opez-Urrutia}},
  \bibnamefont{and} \bibinfo{author}{\bibfnamefont{A.}~\bibnamefont{P\'alffy}},
  \bibinfo{journal}{Phys. Rev. Lett.} \textbf{\bibinfo{volume}{124}},
  \bibinfo{pages}{192502} (\bibinfo{year}{2020}).

\bibitem[{\citenamefont{Dzyublik}(2022)}]{DzyublikPRC2022}
\bibinfo{author}{\bibfnamefont{A.~Y.} \bibnamefont{Dzyublik}},
  \bibinfo{journal}{Phys. Rev. C} \textbf{\bibinfo{volume}{106}},
  \bibinfo{pages}{064608} (\bibinfo{year}{2022}).

\bibitem[{\citenamefont{Karpeshin et~al.}(1996)\citenamefont{Karpeshin, Band,
  Trzhaskovskaya, and Listengarten}}]{KarpeshinPLB1996}
\bibinfo{author}{\bibfnamefont{F.~F.} \bibnamefont{Karpeshin}},
  \bibinfo{author}{\bibfnamefont{I.~M.} \bibnamefont{Band}},
  \bibinfo{author}{\bibfnamefont{M.~B.} \bibnamefont{Trzhaskovskaya}},
  \bibnamefont{and} \bibinfo{author}{\bibfnamefont{M.~A.}
  \bibnamefont{Listengarten}}, \bibinfo{journal}{Phys. Lett. B}
  \textbf{\bibinfo{volume}{372}}, \bibinfo{pages}{1} (\bibinfo{year}{1996}).

\bibitem[{\citenamefont{Karpeshin and Trzhaskovskaya}(2017)}]{KarpeshinPRC2017}
\bibinfo{author}{\bibfnamefont{F.~F.} \bibnamefont{Karpeshin}}
  \bibnamefont{and} \bibinfo{author}{\bibfnamefont{M.~B.}
  \bibnamefont{Trzhaskovskaya}}, \bibinfo{journal}{Phys. Rev. C}
  \textbf{\bibinfo{volume}{95}}, \bibinfo{pages}{034310}
  (\bibinfo{year}{2017}).

\bibitem[{\citenamefont{Xu et~al.}(2023)\citenamefont{Xu, Tang, Wang, Li, Li,
  Cappellaro, and Li}}]{XuPRA2023}
\bibinfo{author}{\bibfnamefont{H.}~\bibnamefont{Xu}},
  \bibinfo{author}{\bibfnamefont{H.}~\bibnamefont{Tang}},
  \bibinfo{author}{\bibfnamefont{G.}~\bibnamefont{Wang}},
  \bibinfo{author}{\bibfnamefont{C.}~\bibnamefont{Li}},
  \bibinfo{author}{\bibfnamefont{B.}~\bibnamefont{Li}},
  \bibinfo{author}{\bibfnamefont{P.}~\bibnamefont{Cappellaro}},
  \bibnamefont{and} \bibinfo{author}{\bibfnamefont{J.}~\bibnamefont{Li}},
  \bibinfo{journal}{Phys. Rev. A} \textbf{\bibinfo{volume}{108}},
  \bibinfo{pages}{L021502} (\bibinfo{year}{2023}).

\bibitem[{\citenamefont{Goldanskii and Namiot}(1976)}]{GoldanskiiPLB1976}
\bibinfo{author}{\bibfnamefont{V.~I.} \bibnamefont{Goldanskii}}
  \bibnamefont{and} \bibinfo{author}{\bibfnamefont{V.~A.}
  \bibnamefont{Namiot}}, \bibinfo{journal}{Phys. Lett. B}
  \textbf{\bibinfo{volume}{62}}, \bibinfo{pages}{393} (\bibinfo{year}{1976}).

\bibitem[{\citenamefont{P\'alffy}(2010)}]{PalffyCP2010}
\bibinfo{author}{\bibfnamefont{A.}~\bibnamefont{P\'alffy}},
  \bibinfo{journal}{Contemporary Physics} \textbf{\bibinfo{volume}{51}},
  \bibinfo{pages}{471} (\bibinfo{year}{2010}).

\bibitem[{\citenamefont{Cue et~al.}(1989)\citenamefont{Cue, Poizat, and
  Remillieux}}]{Cue1989}
\bibinfo{author}{\bibfnamefont{N.}~\bibnamefont{Cue}},
  \bibinfo{author}{\bibfnamefont{J.-C.} \bibnamefont{Poizat}},
  \bibnamefont{and}
  \bibinfo{author}{\bibfnamefont{J.}~\bibnamefont{Remillieux}},
  \bibinfo{journal}{Eurphys.\ Lett.} \textbf{\bibinfo{volume}{8}},
  \bibinfo{pages}{19} (\bibinfo{year}{1989}).

\bibitem[{\citenamefont{Kimball et~al.}(1991)\citenamefont{Kimball, Bittle, and
  Cue}}]{Kimball1}
\bibinfo{author}{\bibfnamefont{J.}~\bibnamefont{Kimball}},
  \bibinfo{author}{\bibfnamefont{D.}~\bibnamefont{Bittle}}, \bibnamefont{and}
  \bibinfo{author}{\bibfnamefont{N.}~\bibnamefont{Cue}},
  \bibinfo{journal}{Phys.\ Lett. A} \textbf{\bibinfo{volume}{152}},
  \bibinfo{pages}{367} (\bibinfo{year}{1991}).

\bibitem[{\citenamefont{Yuan and Kimball}(1993)}]{Kimball2}
\bibinfo{author}{\bibfnamefont{Z.-S.} \bibnamefont{Yuan}} \bibnamefont{and}
  \bibinfo{author}{\bibfnamefont{J.}~\bibnamefont{Kimball}},
  \bibinfo{journal}{Phys.\ Rev.\ C} \textbf{\bibinfo{volume}{47}},
  \bibinfo{pages}{323} (\bibinfo{year}{1993}).

\bibitem[{\citenamefont{Harston and Chemin}(1999)}]{HarstonPRC1999}
\bibinfo{author}{\bibfnamefont{M.~R.} \bibnamefont{Harston}} \bibnamefont{and}
  \bibinfo{author}{\bibfnamefont{J.~F.} \bibnamefont{Chemin}},
  \bibinfo{journal}{Phys. Rev. C} \textbf{\bibinfo{volume}{59}},
  \bibinfo{pages}{2462} (\bibinfo{year}{1999}).

\bibitem[{\citenamefont{Gosselin and Morel}(2004)}]{GosselinPRC2004}
\bibinfo{author}{\bibfnamefont{G.}~\bibnamefont{Gosselin}} \bibnamefont{and}
  \bibinfo{author}{\bibfnamefont{P.}~\bibnamefont{Morel}},
  \bibinfo{journal}{Phys. Rev. C} \textbf{\bibinfo{volume}{70}},
  \bibinfo{pages}{064603} (\bibinfo{year}{2004}).

\bibitem[{\citenamefont{Gosselin et~al.}(2007)\citenamefont{Gosselin, M\'eot,
  and Morel}}]{GosselinPRC2007}
\bibinfo{author}{\bibfnamefont{G.}~\bibnamefont{Gosselin}},
  \bibinfo{author}{\bibfnamefont{V.}~\bibnamefont{M\'eot}}, \bibnamefont{and}
  \bibinfo{author}{\bibfnamefont{P.}~\bibnamefont{Morel}},
  \bibinfo{journal}{Phys. Rev. C} \textbf{\bibinfo{volume}{76}},
  \bibinfo{pages}{044611} (\bibinfo{year}{2007}).

\bibitem[{\citenamefont{Gunst et~al.}(2014)\citenamefont{Gunst, Litvinov,
  Keitel, and P\'alffy}}]{GunstPRL2014}
\bibinfo{author}{\bibfnamefont{J.}~\bibnamefont{Gunst}},
  \bibinfo{author}{\bibfnamefont{Y.~A.} \bibnamefont{Litvinov}},
  \bibinfo{author}{\bibfnamefont{C.~H.} \bibnamefont{Keitel}},
  \bibnamefont{and} \bibinfo{author}{\bibfnamefont{A.}~\bibnamefont{P\'alffy}},
  \bibinfo{journal}{Phys. Rev. Lett.} \textbf{\bibinfo{volume}{112}},
  \bibinfo{pages}{082501} (\bibinfo{year}{2014}).

\bibitem[{\citenamefont{Gunst et~al.}(2015)\citenamefont{Gunst, Wu, Kumar,
  Keitel, and P\'alffy}}]{GunstPOP2015}
\bibinfo{author}{\bibfnamefont{J.}~\bibnamefont{Gunst}},
  \bibinfo{author}{\bibfnamefont{Y.}~\bibnamefont{Wu}},
  \bibinfo{author}{\bibfnamefont{N.}~\bibnamefont{Kumar}},
  \bibinfo{author}{\bibfnamefont{C.~H.} \bibnamefont{Keitel}},
  \bibnamefont{and} \bibinfo{author}{\bibfnamefont{A.}~\bibnamefont{P\'alffy}},
  \bibinfo{journal}{Physics of Plasmas} \textbf{\bibinfo{volume}{22}},
  \bibinfo{pages}{112706} (\bibinfo{year}{2015}).

\bibitem[{\citenamefont{Wu et~al.}(2018)\citenamefont{Wu, Gunst, Keitel, and
  P\'alffy}}]{WuPRL2018}
\bibinfo{author}{\bibfnamefont{Y.}~\bibnamefont{Wu}},
  \bibinfo{author}{\bibfnamefont{J.}~\bibnamefont{Gunst}},
  \bibinfo{author}{\bibfnamefont{C.~H.} \bibnamefont{Keitel}},
  \bibnamefont{and} \bibinfo{author}{\bibfnamefont{A.}~\bibnamefont{P\'alffy}},
  \bibinfo{journal}{Phys. Rev. Lett.} \textbf{\bibinfo{volume}{120}},
  \bibinfo{pages}{052504} (\bibinfo{year}{2018}).

\bibitem[{\citenamefont{Spohr et~al.}(2023)\citenamefont{Spohr, Doria, Baran,
  Cernaianu, Ghenuche, Nastasa, O'Donnell, S\"oderstr\"om, Tudor, Ur
  et~al.}}]{SpohrEPJA2023}
\bibinfo{author}{\bibfnamefont{K.~M.} \bibnamefont{Spohr}},
  \bibinfo{author}{\bibfnamefont{D.}~\bibnamefont{Doria}},
  \bibinfo{author}{\bibfnamefont{V.}~\bibnamefont{Baran}},
  \bibinfo{author}{\bibfnamefont{M.~O.} \bibnamefont{Cernaianu}},
  \bibinfo{author}{\bibfnamefont{P.~V.} \bibnamefont{Ghenuche}},
  \bibinfo{author}{\bibfnamefont{V.}~\bibnamefont{Nastasa}},
  \bibinfo{author}{\bibfnamefont{D.}~\bibnamefont{O'Donnell}},
  \bibinfo{author}{\bibfnamefont{P.-A.} \bibnamefont{S\"oderstr\"om}},
  \bibinfo{author}{\bibfnamefont{L.}~\bibnamefont{Tudor}},
  \bibinfo{author}{\bibfnamefont{C.~A.} \bibnamefont{Ur}},
  \bibnamefont{et~al.}, \bibinfo{journal}{Eur. Phys. J. A}
  \textbf{\bibinfo{volume}{59}}, \bibinfo{pages}{281} (\bibinfo{year}{2023}).

\bibitem[{\citenamefont{Negoita et~al.}(2016)\citenamefont{Negoita, Roth,
  Thirolf, Tudisco, Hannachi, Moustaizis, Pomerantz, Fuchs, Sphor, Acbas
  et~al.}}]{NegoitaRRP2016}
\bibinfo{author}{\bibfnamefont{F.}~\bibnamefont{Negoita}},
  \bibinfo{author}{\bibfnamefont{M.}~\bibnamefont{Roth}},
  \bibinfo{author}{\bibfnamefont{P.~G.} \bibnamefont{Thirolf}},
  \bibinfo{author}{\bibfnamefont{S.}~\bibnamefont{Tudisco}},
  \bibinfo{author}{\bibfnamefont{F.}~\bibnamefont{Hannachi}},
  \bibinfo{author}{\bibfnamefont{S.}~\bibnamefont{Moustaizis}},
  \bibinfo{author}{\bibfnamefont{P.}~\bibnamefont{Pomerantz},
  \bibfnamefont{I~.and~Mckenna}},
  \bibinfo{author}{\bibfnamefont{J.}~\bibnamefont{Fuchs}},
  \bibinfo{author}{\bibfnamefont{K.}~\bibnamefont{Sphor}},
  \bibinfo{author}{\bibfnamefont{G.}~\bibnamefont{Acbas}},
  \bibnamefont{et~al.}, \bibinfo{journal}{Romanian Reports in Physics}
  \textbf{\bibinfo{volume}{68, Supplement}}, \bibinfo{pages}{S37}
  (\bibinfo{year}{2016}).

\bibitem[{\citenamefont{Cerjan et~al.}(2018)\citenamefont{Cerjan, Bernstein,
  Hopkins, Bionta, Bleuel, Caggiano, Cassata, Brune, Frenje, Gatu-Johnson
  et~al.}}]{NIF2018}
\bibinfo{author}{\bibfnamefont{C.~J.} \bibnamefont{Cerjan}},
  \bibinfo{author}{\bibfnamefont{L.}~\bibnamefont{Bernstein}},
  \bibinfo{author}{\bibfnamefont{L.~B.} \bibnamefont{Hopkins}},
  \bibinfo{author}{\bibfnamefont{R.~M.} \bibnamefont{Bionta}},
  \bibinfo{author}{\bibfnamefont{D.~L.} \bibnamefont{Bleuel}},
  \bibinfo{author}{\bibfnamefont{J.~A.} \bibnamefont{Caggiano}},
  \bibinfo{author}{\bibfnamefont{W.~S.} \bibnamefont{Cassata}},
  \bibinfo{author}{\bibfnamefont{C.~R.} \bibnamefont{Brune}},
  \bibinfo{author}{\bibfnamefont{J.}~\bibnamefont{Frenje}},
  \bibinfo{author}{\bibfnamefont{M.}~\bibnamefont{Gatu-Johnson}},
  \bibnamefont{et~al.}, \bibinfo{journal}{Journal of Physics G: Nuclear and
  Particle Physics} \textbf{\bibinfo{volume}{45}}, \bibinfo{pages}{033003}
  (\bibinfo{year}{2018}).

\bibitem[{\citenamefont{Doolen}(1978{\natexlab{a}})}]{doolen1978nuclear}
\bibinfo{author}{\bibfnamefont{G.~D.} \bibnamefont{Doolen}},
  \bibinfo{journal}{Phys. Rev. Lett.} \textbf{\bibinfo{volume}{40}},
  \bibinfo{pages}{1695} (\bibinfo{year}{1978}{\natexlab{a}}).

\bibitem[{\citenamefont{Doolen}(1978{\natexlab{b}})}]{doolen1978nuclearC}
\bibinfo{author}{\bibfnamefont{G.~D.} \bibnamefont{Doolen}},
  \bibinfo{journal}{Phys. Rev. C} \textbf{\bibinfo{volume}{18}},
  \bibinfo{pages}{2547} (\bibinfo{year}{1978}{\natexlab{b}}).

\bibitem[{\citenamefont{P\'alffy et~al.}(2006)\citenamefont{P\'alffy, Scheid,
  and Harman}}]{PalffyPRA2006}
\bibinfo{author}{\bibfnamefont{A.}~\bibnamefont{P\'alffy}},
  \bibinfo{author}{\bibfnamefont{W.}~\bibnamefont{Scheid}}, \bibnamefont{and}
  \bibinfo{author}{\bibfnamefont{Z.}~\bibnamefont{Harman}},
  \bibinfo{journal}{Phys. Rev. A} \textbf{\bibinfo{volume}{73}},
  \bibinfo{pages}{012715} (\bibinfo{year}{2006}).

\bibitem[{\citenamefont{P\'alffy et~al.}(2008)\citenamefont{P\'alffy, Harman,
  Kozhuharov, Brandau, Keitel, Scheid, and St\"ohlker}}]{PalffyPLB2008}
\bibinfo{author}{\bibfnamefont{A.}~\bibnamefont{P\'alffy}},
  \bibinfo{author}{\bibfnamefont{Z.}~\bibnamefont{Harman}},
  \bibinfo{author}{\bibfnamefont{C.}~\bibnamefont{Kozhuharov}},
  \bibinfo{author}{\bibfnamefont{C.}~\bibnamefont{Brandau}},
  \bibinfo{author}{\bibfnamefont{C.~H.} \bibnamefont{Keitel}},
  \bibinfo{author}{\bibfnamefont{W.}~\bibnamefont{Scheid}}, \bibnamefont{and}
  \bibinfo{author}{\bibfnamefont{T.}~\bibnamefont{St\"ohlker}},
  \bibinfo{journal}{Phys. Lett. B} \textbf{\bibinfo{volume}{661}},
  \bibinfo{pages}{330} (\bibinfo{year}{2008}).

\bibitem[{\citenamefont{Ma et~al.}(2015)\citenamefont{Ma, Wen, Huang, Wang,
  Yuan, Wang, Sun, Mao, Yang, Xu et~al.}}]{MaPS2015}
\bibinfo{author}{\bibfnamefont{X.}~\bibnamefont{Ma}},
  \bibinfo{author}{\bibfnamefont{W.~Q.} \bibnamefont{Wen}},
  \bibinfo{author}{\bibfnamefont{Z.~K.} \bibnamefont{Huang}},
  \bibinfo{author}{\bibfnamefont{H.~B.} \bibnamefont{Wang}},
  \bibinfo{author}{\bibfnamefont{Y.~J.} \bibnamefont{Yuan}},
  \bibinfo{author}{\bibfnamefont{M.}~\bibnamefont{Wang}},
  \bibinfo{author}{\bibfnamefont{Z.~Y.} \bibnamefont{Sun}},
  \bibinfo{author}{\bibfnamefont{L.~J.} \bibnamefont{Mao}},
  \bibinfo{author}{\bibfnamefont{J.~C.} \bibnamefont{Yang}},
  \bibinfo{author}{\bibfnamefont{H.~S.} \bibnamefont{Xu}},
  \bibnamefont{et~al.}, \bibinfo{journal}{Phys. Scr.}
  \textbf{\bibinfo{volume}{T166}}, \bibinfo{pages}{014012}
  (\bibinfo{year}{2015}).

\bibitem[{\citenamefont{Yang et~al.}(2024)\citenamefont{Yang, Wang, Ma, Fu, He,
  and Ma}}]{YangFP2024}
\bibinfo{author}{\bibfnamefont{Y.}~\bibnamefont{Yang}},
  \bibinfo{author}{\bibfnamefont{Y.}~\bibnamefont{Wang}},
  \bibinfo{author}{\bibfnamefont{Z.}~\bibnamefont{Ma}},
  \bibinfo{author}{\bibfnamefont{C.}~\bibnamefont{Fu}},
  \bibinfo{author}{\bibfnamefont{W.}~\bibnamefont{He}}, \bibnamefont{and}
  \bibinfo{author}{\bibfnamefont{Y.}~\bibnamefont{Ma}},
  \bibinfo{journal}{Front. Phys., accepted [doi: 10.3389/fphy.2024.1410076]}
  (\bibinfo{year}{2024}).

\bibitem[{\citenamefont{P\'alffy
  et~al.}(2007{\natexlab{a}})\citenamefont{P\'alffy, Harman, and
  Scheid}}]{PalffyPRA2007}
\bibinfo{author}{\bibfnamefont{A.}~\bibnamefont{P\'alffy}},
  \bibinfo{author}{\bibfnamefont{Z.}~\bibnamefont{Harman}}, \bibnamefont{and}
  \bibinfo{author}{\bibfnamefont{W.}~\bibnamefont{Scheid}},
  \bibinfo{journal}{Phys. Rev. A} \textbf{\bibinfo{volume}{75}},
  \bibinfo{pages}{012709} (\bibinfo{year}{2007}{\natexlab{a}}).

\bibitem[{\citenamefont{Dillmann}(2022)}]{DillmannBerlin2022}
\bibinfo{author}{\bibfnamefont{I.}~\bibnamefont{Dillmann}},
  \emph{\bibinfo{title}{New Experimental Approaches to Nuclear Isomers for
  Nuclear Astrophysics}} (\bibinfo{publisher}{100 Years of Nuclear Isomers,
  workshop}, \bibinfo{address}{Berlin-Dahlem}, \bibinfo{year}{2022}).

\bibitem[{\citenamefont{Wang et~al.}(2023)\citenamefont{Wang, Ma, Yang, Fu, He,
  and Ma}}]{WangFP2023}
\bibinfo{author}{\bibfnamefont{Y.}~\bibnamefont{Wang}},
  \bibinfo{author}{\bibfnamefont{Z.}~\bibnamefont{Ma}},
  \bibinfo{author}{\bibfnamefont{Y.}~\bibnamefont{Yang}},
  \bibinfo{author}{\bibfnamefont{C.}~\bibnamefont{Fu}},
  \bibinfo{author}{\bibfnamefont{W.}~\bibnamefont{He}}, \bibnamefont{and}
  \bibinfo{author}{\bibfnamefont{Y.}~\bibnamefont{Ma}},
  \bibinfo{journal}{Front. Phys.} \textbf{\bibinfo{volume}{11}},
  \bibinfo{pages}{1203401} (\bibinfo{year}{2023}).

\bibitem[{\citenamefont{Madan et~al.}(2020)\citenamefont{Madan, Vanacore,
  Gargiulo, LaGrange, and Carbone}}]{Madan2020}
\bibinfo{author}{\bibfnamefont{I.}~\bibnamefont{Madan}},
  \bibinfo{author}{\bibfnamefont{G.~M.} \bibnamefont{Vanacore}},
  \bibinfo{author}{\bibfnamefont{S.}~\bibnamefont{Gargiulo}},
  \bibinfo{author}{\bibfnamefont{T.}~\bibnamefont{LaGrange}}, \bibnamefont{and}
  \bibinfo{author}{\bibfnamefont{F.}~\bibnamefont{Carbone}},
  \bibinfo{journal}{Applied Physics Letters} \textbf{\bibinfo{volume}{116}},
  \bibinfo{pages}{230502} (\bibinfo{year}{2020}).

\bibitem[{\citenamefont{Wu et~al.}(2022)\citenamefont{Wu, Gargiulo, Carbone,
  Keitel, and P\'alffy}}]{WuPRL2022}
\bibinfo{author}{\bibfnamefont{Y.}~\bibnamefont{Wu}},
  \bibinfo{author}{\bibfnamefont{S.}~\bibnamefont{Gargiulo}},
  \bibinfo{author}{\bibfnamefont{F.}~\bibnamefont{Carbone}},
  \bibinfo{author}{\bibfnamefont{C.~H.} \bibnamefont{Keitel}},
  \bibnamefont{and} \bibinfo{author}{\bibfnamefont{A.}~\bibnamefont{P\'alffy}},
  \bibinfo{journal}{Phys. Rev. Lett.} \textbf{\bibinfo{volume}{128}},
  \bibinfo{pages}{162501} (\bibinfo{year}{2022}).

\bibitem[{\citenamefont{Gargiulo et~al.}(2022)\citenamefont{Gargiulo, Madan,
  and Carbone}}]{GargiuloPRL2022}
\bibinfo{author}{\bibfnamefont{S.}~\bibnamefont{Gargiulo}},
  \bibinfo{author}{\bibfnamefont{I.}~\bibnamefont{Madan}}, \bibnamefont{and}
  \bibinfo{author}{\bibfnamefont{F.}~\bibnamefont{Carbone}},
  \bibinfo{journal}{Phys. Rev. Lett.} \textbf{\bibinfo{volume}{128}},
  \bibinfo{pages}{212502} (\bibinfo{year}{2022}).

\bibitem[{\citenamefont{Chiara et~al.}(2018)\citenamefont{Chiara, Carroll,
  Carpenter, Greene, Hartley, Janssens, Lane, Marsh, Matters, Polasik
  et~al.}}]{ChiaraNature2018}
\bibinfo{author}{\bibfnamefont{C.~J.} \bibnamefont{Chiara}},
  \bibinfo{author}{\bibfnamefont{J.~J.} \bibnamefont{Carroll}},
  \bibinfo{author}{\bibfnamefont{M.~P.} \bibnamefont{Carpenter}},
  \bibinfo{author}{\bibfnamefont{J.~P.} \bibnamefont{Greene}},
  \bibinfo{author}{\bibfnamefont{D.~J.} \bibnamefont{Hartley}},
  \bibinfo{author}{\bibfnamefont{R.~V.~F.} \bibnamefont{Janssens}},
  \bibinfo{author}{\bibfnamefont{G.~J.} \bibnamefont{Lane}},
  \bibinfo{author}{\bibfnamefont{J.~C.} \bibnamefont{Marsh}},
  \bibinfo{author}{\bibfnamefont{D.~A.} \bibnamefont{Matters}},
  \bibinfo{author}{\bibfnamefont{M.}~\bibnamefont{Polasik}},
  \bibnamefont{et~al.}, \bibinfo{journal}{Nature}
  \textbf{\bibinfo{volume}{554}}, \bibinfo{pages}{216} (\bibinfo{year}{2018}).

\bibitem[{\citenamefont{Wu et~al.}(2019)\citenamefont{Wu, Keitel, and
  P\'alffy}}]{Wu2019}
\bibinfo{author}{\bibfnamefont{Y.}~\bibnamefont{Wu}},
  \bibinfo{author}{\bibfnamefont{C.~H.} \bibnamefont{Keitel}},
  \bibnamefont{and} \bibinfo{author}{\bibfnamefont{A.}~\bibnamefont{P\'alffy}},
  \bibinfo{journal}{Phys. Rev. Lett.} \textbf{\bibinfo{volume}{122}},
  \bibinfo{pages}{212501} (\bibinfo{year}{2019}).

\bibitem[{\citenamefont{Rzadkiewicz et~al.}(2021)\citenamefont{Rzadkiewicz,
  Polasik, S\l{}abkowska, Syrocki, Carroll, and Chiara}}]{Polasik2021}
\bibinfo{author}{\bibfnamefont{J.}~\bibnamefont{Rzadkiewicz}},
  \bibinfo{author}{\bibfnamefont{M.}~\bibnamefont{Polasik}},
  \bibinfo{author}{\bibfnamefont{K.}~\bibnamefont{S\l{}abkowska}},
  \bibinfo{author}{\bibfnamefont{L.}~\bibnamefont{Syrocki}},
  \bibinfo{author}{\bibfnamefont{J.~J.} \bibnamefont{Carroll}},
  \bibnamefont{and} \bibinfo{author}{\bibfnamefont{C.~J.}
  \bibnamefont{Chiara}}, \bibinfo{journal}{Phys. Rev. Lett.}
  \textbf{\bibinfo{volume}{127}}, \bibinfo{pages}{042501}
  (\bibinfo{year}{2021}).

\bibitem[{\citenamefont{Rzadkiewicz et~al.}(2023)\citenamefont{Rzadkiewicz,
  S\l{}abkowska, Polasik, Syrocki, Carroll, and Chiara}}]{RzadkiewiczPRC2023}
\bibinfo{author}{\bibfnamefont{J.}~\bibnamefont{Rzadkiewicz}},
  \bibinfo{author}{\bibfnamefont{K.}~\bibnamefont{S\l{}abkowska}},
  \bibinfo{author}{\bibfnamefont{M.}~\bibnamefont{Polasik}},
  \bibinfo{author}{\bibfnamefont{L.}~\bibnamefont{Syrocki}},
  \bibinfo{author}{\bibfnamefont{J.~J.} \bibnamefont{Carroll}},
  \bibnamefont{and} \bibinfo{author}{\bibfnamefont{C.~J.}
  \bibnamefont{Chiara}}, \bibinfo{journal}{Phys. Rev. C}
  \textbf{\bibinfo{volume}{108}}, \bibinfo{pages}{L031302}
  (\bibinfo{year}{2023}).

\bibitem[{\citenamefont{Guo et~al.}(2021)\citenamefont{Guo, Fang, Zhou, and
  Petrache}}]{Guo2021}
\bibinfo{author}{\bibfnamefont{S.}~\bibnamefont{Guo}},
  \bibinfo{author}{\bibfnamefont{Y.}~\bibnamefont{Fang}},
  \bibinfo{author}{\bibfnamefont{X.}~\bibnamefont{Zhou}}, \bibnamefont{and}
  \bibinfo{author}{\bibfnamefont{C.~M.} \bibnamefont{Petrache}},
  \bibinfo{journal}{Nature} \textbf{\bibinfo{volume}{594}}, \bibinfo{pages}{E1}
  (\bibinfo{year}{2021}).

\bibitem[{\citenamefont{Chiara et~al.}(2021)\citenamefont{Chiara, Carroll,
  Carpenter, Greene, Hartley, Janssens, Lane, Marsh, Matters, Polasik
  et~al.}}]{Chiara2021}
\bibinfo{author}{\bibfnamefont{C.~J.} \bibnamefont{Chiara}},
  \bibinfo{author}{\bibfnamefont{J.~J.} \bibnamefont{Carroll}},
  \bibinfo{author}{\bibfnamefont{M.~P.} \bibnamefont{Carpenter}},
  \bibinfo{author}{\bibfnamefont{J.~P.} \bibnamefont{Greene}},
  \bibinfo{author}{\bibfnamefont{D.~J.} \bibnamefont{Hartley}},
  \bibinfo{author}{\bibfnamefont{R.~V.~F.} \bibnamefont{Janssens}},
  \bibinfo{author}{\bibfnamefont{G.~J.} \bibnamefont{Lane}},
  \bibinfo{author}{\bibfnamefont{J.~C.} \bibnamefont{Marsh}},
  \bibinfo{author}{\bibfnamefont{D.~A.} \bibnamefont{Matters}},
  \bibinfo{author}{\bibfnamefont{M.}~\bibnamefont{Polasik}},
  \bibnamefont{et~al.}, \bibinfo{journal}{Nature}
  \textbf{\bibinfo{volume}{594}}, \bibinfo{pages}{E3} (\bibinfo{year}{2021}).

\bibitem[{\citenamefont{Guo et~al.}(2022)\citenamefont{Guo, Ding, Zhou, Wu,
  Wang, Xu, Fang, Petrache, Lawrie, Qiang et~al.}}]{GuoPRL2022}
\bibinfo{author}{\bibfnamefont{S.}~\bibnamefont{Guo}},
  \bibinfo{author}{\bibfnamefont{B.}~\bibnamefont{Ding}},
  \bibinfo{author}{\bibfnamefont{X.~H.} \bibnamefont{Zhou}},
  \bibinfo{author}{\bibfnamefont{Y.~B.} \bibnamefont{Wu}},
  \bibinfo{author}{\bibfnamefont{J.~G.} \bibnamefont{Wang}},
  \bibinfo{author}{\bibfnamefont{S.~W.} \bibnamefont{Xu}},
  \bibinfo{author}{\bibfnamefont{Y.~D.} \bibnamefont{Fang}},
  \bibinfo{author}{\bibfnamefont{C.~M.} \bibnamefont{Petrache}},
  \bibinfo{author}{\bibfnamefont{E.~A.} \bibnamefont{Lawrie}},
  \bibinfo{author}{\bibfnamefont{Y.~H.} \bibnamefont{Qiang}},
  \bibnamefont{et~al.}, \bibinfo{journal}{Phys. Rev. Lett.}
  \textbf{\bibinfo{volume}{128}}, \bibinfo{pages}{242502}
  (\bibinfo{year}{2022}).

\bibitem[{\citenamefont{Donets}(1990)}]{DonetsRSI1990}
\bibinfo{author}{\bibfnamefont{E.~D.} \bibnamefont{Donets}},
  \bibinfo{journal}{Rev. Sci. Instrum.} \textbf{\bibinfo{volume}{61}},
  \bibinfo{pages}{225} (\bibinfo{year}{1990}).

\bibitem[{\citenamefont{Marrs et~al.}(1988)\citenamefont{Marrs, Levine, Knapp,
  and Henderson}}]{MarrsPRL1988}
\bibinfo{author}{\bibfnamefont{R.~E.} \bibnamefont{Marrs}},
  \bibinfo{author}{\bibfnamefont{M.~A.} \bibnamefont{Levine}},
  \bibinfo{author}{\bibfnamefont{D.~A.} \bibnamefont{Knapp}}, \bibnamefont{and}
  \bibinfo{author}{\bibfnamefont{J.~R.} \bibnamefont{Henderson}},
  \bibinfo{journal}{Phys. Rev. Lett.} \textbf{\bibinfo{volume}{60}},
  \bibinfo{pages}{1715} (\bibinfo{year}{1988}).

\bibitem[{\citenamefont{Levine et~al.}(1988)\citenamefont{Levine, Marrs,
  Henderson, Knapp, and Schneider}}]{LevinePS1988}
\bibinfo{author}{\bibfnamefont{M.~A.} \bibnamefont{Levine}},
  \bibinfo{author}{\bibfnamefont{R.~E.} \bibnamefont{Marrs}},
  \bibinfo{author}{\bibfnamefont{J.~R.} \bibnamefont{Henderson}},
  \bibinfo{author}{\bibfnamefont{D.~A.} \bibnamefont{Knapp}}, \bibnamefont{and}
  \bibinfo{author}{\bibfnamefont{M.~B.} \bibnamefont{Schneider}},
  \bibinfo{journal}{Phys. Scr.} \textbf{\bibinfo{volume}{T22}},
  \bibinfo{pages}{157} (\bibinfo{year}{1988}).

\bibitem[{\citenamefont{Gillaspy}(2001)}]{GillaspyJPB2001}
\bibinfo{author}{\bibfnamefont{J.~D.} \bibnamefont{Gillaspy}},
  \bibinfo{journal}{J. Phys. B: At. Mol. Opt. Phys.}
  \textbf{\bibinfo{volume}{34}}, \bibinfo{pages}{R93} (\bibinfo{year}{2001}).

\bibitem[{\citenamefont{Currell and Fussmann}(2005)}]{CurrellIEEE2005}
\bibinfo{author}{\bibfnamefont{F.}~\bibnamefont{Currell}} \bibnamefont{and}
  \bibinfo{author}{\bibfnamefont{G.}~\bibnamefont{Fussmann}},
  \bibinfo{journal}{IEEE Transactions of Plasma Science}
  \textbf{\bibinfo{volume}{33}}, \bibinfo{pages}{1763} (\bibinfo{year}{2005}).

\bibitem[{\citenamefont{Blessenohl et~al.}(2018)\citenamefont{Blessenohl,
  Dobrodey, Warnecke, Rosner, Graham, Paul, Baumann, Hockenbery, Hubele,
  Pfeifer et~al.}}]{BlessenohlRSI2018}
\bibinfo{author}{\bibfnamefont{M.~A.} \bibnamefont{Blessenohl}},
  \bibinfo{author}{\bibfnamefont{S.}~\bibnamefont{Dobrodey}},
  \bibinfo{author}{\bibfnamefont{C.}~\bibnamefont{Warnecke}},
  \bibinfo{author}{\bibfnamefont{M.~K.} \bibnamefont{Rosner}},
  \bibinfo{author}{\bibfnamefont{L.}~\bibnamefont{Graham}},
  \bibinfo{author}{\bibfnamefont{S.}~\bibnamefont{Paul}},
  \bibinfo{author}{\bibfnamefont{T.~M.} \bibnamefont{Baumann}},
  \bibinfo{author}{\bibfnamefont{Z.}~\bibnamefont{Hockenbery}},
  \bibinfo{author}{\bibfnamefont{R.}~\bibnamefont{Hubele}},
  \bibinfo{author}{\bibfnamefont{T.}~\bibnamefont{Pfeifer}},
  \bibnamefont{et~al.}, \bibinfo{journal}{Rev. Sci. Instrum.}
  \textbf{\bibinfo{volume}{89}}, \bibinfo{pages}{052401}
  (\bibinfo{year}{2018}).

\bibitem[{\citenamefont{Micke et~al.}(2018)\citenamefont{Micke, K\"uhn,
  Buchauer, Harries, B\"ucking, Blaum, Cieluch, Egl, Hollain, Kraemer
  et~al.}}]{MickeRSI2018}
\bibinfo{author}{\bibfnamefont{P.}~\bibnamefont{Micke}},
  \bibinfo{author}{\bibfnamefont{S.}~\bibnamefont{K\"uhn}},
  \bibinfo{author}{\bibfnamefont{L.}~\bibnamefont{Buchauer}},
  \bibinfo{author}{\bibfnamefont{J.~R.} \bibnamefont{Harries}},
  \bibinfo{author}{\bibfnamefont{T.~M.} \bibnamefont{B\"ucking}},
  \bibinfo{author}{\bibfnamefont{K.}~\bibnamefont{Blaum}},
  \bibinfo{author}{\bibfnamefont{A.}~\bibnamefont{Cieluch}},
  \bibinfo{author}{\bibfnamefont{A.}~\bibnamefont{Egl}},
  \bibinfo{author}{\bibfnamefont{D.}~\bibnamefont{Hollain}},
  \bibinfo{author}{\bibfnamefont{S.}~\bibnamefont{Kraemer}},
  \bibnamefont{et~al.}, \bibinfo{journal}{Rev. Sci. Instrum.}
  \textbf{\bibinfo{volume}{89}}, \bibinfo{pages}{063109}
  (\bibinfo{year}{2018}).

\bibitem[{\citenamefont{Schweiger et~al.}(2019)\citenamefont{Schweiger,
  K\"onig, Crespo L\'opez-Urrutia, Door, Dorrer, D\"ullmann, Eliseev, Filianin,
  Huang, Kromer et~al.}}]{SchweigerRSI2019}
\bibinfo{author}{\bibfnamefont{C.}~\bibnamefont{Schweiger}},
  \bibinfo{author}{\bibfnamefont{C.~M.} \bibnamefont{K\"onig}},
  \bibinfo{author}{\bibfnamefont{J.~R.} \bibnamefont{Crespo L\'opez-Urrutia}},
  \bibinfo{author}{\bibfnamefont{M.}~\bibnamefont{Door}},
  \bibinfo{author}{\bibfnamefont{H.}~\bibnamefont{Dorrer}},
  \bibinfo{author}{\bibfnamefont{C.~E.} \bibnamefont{D\"ullmann}},
  \bibinfo{author}{\bibfnamefont{S.}~\bibnamefont{Eliseev}},
  \bibinfo{author}{\bibfnamefont{P.}~\bibnamefont{Filianin}},
  \bibinfo{author}{\bibfnamefont{W.}~\bibnamefont{Huang}},
  \bibinfo{author}{\bibfnamefont{K.}~\bibnamefont{Kromer}},
  \bibnamefont{et~al.}, \bibinfo{journal}{Rev. Sci. Instrum.}
  \textbf{\bibinfo{volume}{90}}, \bibinfo{pages}{123201}
  (\bibinfo{year}{2019}).

\bibitem[{\citenamefont{Blaum}(2006)}]{BlaumPR2006}
\bibinfo{author}{\bibfnamefont{K.}~\bibnamefont{Blaum}},
  \bibinfo{journal}{Physics Reports} \textbf{\bibinfo{volume}{425}},
  \bibinfo{pages}{1} (\bibinfo{year}{2006}).

\bibitem[{\citenamefont{Dilling et~al.}(2018)\citenamefont{Dilling, Blaum,
  Brodeur, and Eliseev}}]{DillingARNPS2018}
\bibinfo{author}{\bibfnamefont{J.}~\bibnamefont{Dilling}},
  \bibinfo{author}{\bibfnamefont{K.}~\bibnamefont{Blaum}},
  \bibinfo{author}{\bibfnamefont{M.}~\bibnamefont{Brodeur}}, \bibnamefont{and}
  \bibinfo{author}{\bibfnamefont{S.}~\bibnamefont{Eliseev}},
  \bibinfo{journal}{Annu. Rev. Nucl. Part. Sci.} \textbf{\bibinfo{volume}{68}},
  \bibinfo{pages}{45} (\bibinfo{year}{2018}).

\bibitem[{\citenamefont{Blaum et~al.}(2021)\citenamefont{Blaum, Eliseev, and
  Sturm}}]{BlaumQST2021}
\bibinfo{author}{\bibfnamefont{K.}~\bibnamefont{Blaum}},
  \bibinfo{author}{\bibfnamefont{S.}~\bibnamefont{Eliseev}}, \bibnamefont{and}
  \bibinfo{author}{\bibfnamefont{S.}~\bibnamefont{Sturm}},
  \bibinfo{journal}{Quantum Sci. Technol.} \textbf{\bibinfo{volume}{6}},
  \bibinfo{pages}{014002} (\bibinfo{year}{2021}).

\bibitem[{\citenamefont{Kozlov et~al.}(2018)\citenamefont{Kozlov, Safronova,
  Crespo L\'opez-Urrutia, and Schmidt}}]{KozlovRMP2018}
\bibinfo{author}{\bibfnamefont{M.~G.} \bibnamefont{Kozlov}},
  \bibinfo{author}{\bibfnamefont{M.~S.} \bibnamefont{Safronova}},
  \bibinfo{author}{\bibfnamefont{J.~R.} \bibnamefont{Crespo L\'opez-Urrutia}},
  \bibnamefont{and} \bibinfo{author}{\bibfnamefont{P.~O.}
  \bibnamefont{Schmidt}}, \bibinfo{journal}{Rev. Mod. Phys.}
  \textbf{\bibinfo{volume}{90}}, \bibinfo{pages}{045005}
  (\bibinfo{year}{2018}).

\bibitem[{\citenamefont{Geissel et~al.}(1992)\citenamefont{Geissel, Beckert,
  Bosch, Eickhoff, Franczak, Franzke, Jung, Klepper, Moshammer, M\"unzenberg
  et~al.}}]{GeisselPRL1992}
\bibinfo{author}{\bibfnamefont{H.}~\bibnamefont{Geissel}},
  \bibinfo{author}{\bibfnamefont{K.}~\bibnamefont{Beckert}},
  \bibinfo{author}{\bibfnamefont{F.}~\bibnamefont{Bosch}},
  \bibinfo{author}{\bibfnamefont{H.}~\bibnamefont{Eickhoff}},
  \bibinfo{author}{\bibfnamefont{B.}~\bibnamefont{Franczak}},
  \bibinfo{author}{\bibfnamefont{B.}~\bibnamefont{Franzke}},
  \bibinfo{author}{\bibfnamefont{M.}~\bibnamefont{Jung}},
  \bibinfo{author}{\bibfnamefont{O.}~\bibnamefont{Klepper}},
  \bibinfo{author}{\bibfnamefont{R.}~\bibnamefont{Moshammer}},
  \bibinfo{author}{\bibfnamefont{G.}~\bibnamefont{M\"unzenberg}},
  \bibnamefont{et~al.}, \bibinfo{journal}{Phys. Rev. Lett.}
  \textbf{\bibinfo{volume}{68}}, \bibinfo{pages}{3412} (\bibinfo{year}{1992}).

\bibitem[{\citenamefont{Xia et~al.}(2002)\citenamefont{Xia, Zhan, Wei, Yuan,
  Song, Zhang, Yang, Yuan, Gao, Zhao et~al.}}]{XiaNIMPRA2022}
\bibinfo{author}{\bibfnamefont{J.~W.} \bibnamefont{Xia}},
  \bibinfo{author}{\bibfnamefont{W.~L.} \bibnamefont{Zhan}},
  \bibinfo{author}{\bibfnamefont{B.~W.} \bibnamefont{Wei}},
  \bibinfo{author}{\bibfnamefont{Y.~J.} \bibnamefont{Yuan}},
  \bibinfo{author}{\bibfnamefont{M.~T.} \bibnamefont{Song}},
  \bibinfo{author}{\bibfnamefont{W.~Z.} \bibnamefont{Zhang}},
  \bibinfo{author}{\bibfnamefont{X.~D.} \bibnamefont{Yang}},
  \bibinfo{author}{\bibfnamefont{P.}~\bibnamefont{Yuan}},
  \bibinfo{author}{\bibfnamefont{D.~Q.} \bibnamefont{Gao}},
  \bibinfo{author}{\bibfnamefont{H.~W.} \bibnamefont{Zhao}},
  \bibnamefont{et~al.}, \bibinfo{journal}{Nuclear Instruments and Methods in
  Physics Research A} \textbf{\bibinfo{volume}{488}}, \bibinfo{pages}{11}
  (\bibinfo{year}{2002}).

\bibitem[{\citenamefont{Litvinov and Bosch}(2011)}]{LitvinovRPP2011}
\bibinfo{author}{\bibfnamefont{Y.~A.} \bibnamefont{Litvinov}} \bibnamefont{and}
  \bibinfo{author}{\bibfnamefont{F.}~\bibnamefont{Bosch}},
  \bibinfo{journal}{Rep. Prog. Phys.} \textbf{\bibinfo{volume}{74}},
  \bibinfo{pages}{016301} (\bibinfo{year}{2011}).

\bibitem[{\citenamefont{Schippers et~al.}(2007)\citenamefont{Schippers,
  Schmidt, Bernhardt, Yu, M\"uller, Lestinsky, Orlov, Grieser, Repnow, and
  Wolf}}]{SchippersPRL2007}
\bibinfo{author}{\bibfnamefont{S.}~\bibnamefont{Schippers}},
  \bibinfo{author}{\bibfnamefont{E.~W.} \bibnamefont{Schmidt}},
  \bibinfo{author}{\bibfnamefont{D.}~\bibnamefont{Bernhardt}},
  \bibinfo{author}{\bibfnamefont{D.}~\bibnamefont{Yu}},
  \bibinfo{author}{\bibfnamefont{A.}~\bibnamefont{M\"uller}},
  \bibinfo{author}{\bibfnamefont{M.}~\bibnamefont{Lestinsky}},
  \bibinfo{author}{\bibfnamefont{D.~A.} \bibnamefont{Orlov}},
  \bibinfo{author}{\bibfnamefont{M.}~\bibnamefont{Grieser}},
  \bibinfo{author}{\bibfnamefont{R.}~\bibnamefont{Repnow}}, \bibnamefont{and}
  \bibinfo{author}{\bibfnamefont{A.}~\bibnamefont{Wolf}},
  \bibinfo{journal}{Phys. Rev. Lett.} \textbf{\bibinfo{volume}{98}},
  \bibinfo{pages}{033001} (\bibinfo{year}{2007}).

\bibitem[{\citenamefont{Brandau et~al.}(2013)\citenamefont{Brandau, Kozhuharov,
  M\"uller, Bernhardt, Banas, Bosch, Currell, Dimopoulou, Gumberidze, Hagmann
  et~al.}}]{BrandauPS2013}
\bibinfo{author}{\bibfnamefont{C.}~\bibnamefont{Brandau}},
  \bibinfo{author}{\bibfnamefont{C.}~\bibnamefont{Kozhuharov}},
  \bibinfo{author}{\bibfnamefont{A.}~\bibnamefont{M\"uller}},
  \bibinfo{author}{\bibfnamefont{D.}~\bibnamefont{Bernhardt}},
  \bibinfo{author}{\bibfnamefont{D.}~\bibnamefont{Banas}},
  \bibinfo{author}{\bibfnamefont{F.}~\bibnamefont{Bosch}},
  \bibinfo{author}{\bibfnamefont{F.~J.} \bibnamefont{Currell}},
  \bibinfo{author}{\bibfnamefont{C.}~\bibnamefont{Dimopoulou}},
  \bibinfo{author}{\bibfnamefont{A.}~\bibnamefont{Gumberidze}},
  \bibinfo{author}{\bibfnamefont{S.}~\bibnamefont{Hagmann}},
  \bibnamefont{et~al.}, \bibinfo{journal}{Phys. Scr.}
  \textbf{\bibinfo{volume}{T156}}, \bibinfo{pages}{014050}
  (\bibinfo{year}{2013}).

\bibitem[{\citenamefont{Grieser et~al.}(2012)\citenamefont{Grieser, Litvinov,
  Raabe, Blaum, Blumenfeld, Butler, Wenander, Woods, Aliotta, Andreyev
  et~al.}}]{GrieserEPJST2012}
\bibinfo{author}{\bibfnamefont{M.}~\bibnamefont{Grieser}},
  \bibinfo{author}{\bibfnamefont{Y.~A.} \bibnamefont{Litvinov}},
  \bibinfo{author}{\bibfnamefont{R.}~\bibnamefont{Raabe}},
  \bibinfo{author}{\bibfnamefont{K.}~\bibnamefont{Blaum}},
  \bibinfo{author}{\bibfnamefont{Y.}~\bibnamefont{Blumenfeld}},
  \bibinfo{author}{\bibfnamefont{P.~A.} \bibnamefont{Butler}},
  \bibinfo{author}{\bibfnamefont{F.}~\bibnamefont{Wenander}},
  \bibinfo{author}{\bibfnamefont{P.~J.} \bibnamefont{Woods}},
  \bibinfo{author}{\bibfnamefont{M.}~\bibnamefont{Aliotta}},
  \bibinfo{author}{\bibfnamefont{A.}~\bibnamefont{Andreyev}},
  \bibnamefont{et~al.}, \bibinfo{journal}{European Physical Journal Special
  Topics} \textbf{\bibinfo{volume}{207}}, \bibinfo{pages}{1}
  (\bibinfo{year}{2012}).

\bibitem[{\citenamefont{Litvinov et~al.}(2013)\citenamefont{Litvinov, Bishop,
  Blaum, Bosch, Brandau, Chen, Dillmann, Egelhof, Geissel, Grisenti
  et~al.}}]{LitvinovNIMPRB2013}
\bibinfo{author}{\bibfnamefont{Y.~A.} \bibnamefont{Litvinov}},
  \bibinfo{author}{\bibfnamefont{S.}~\bibnamefont{Bishop}},
  \bibinfo{author}{\bibfnamefont{K.}~\bibnamefont{Blaum}},
  \bibinfo{author}{\bibfnamefont{F.}~\bibnamefont{Bosch}},
  \bibinfo{author}{\bibfnamefont{C.}~\bibnamefont{Brandau}},
  \bibinfo{author}{\bibfnamefont{L.~X.} \bibnamefont{Chen}},
  \bibinfo{author}{\bibfnamefont{I.}~\bibnamefont{Dillmann}},
  \bibinfo{author}{\bibfnamefont{P.}~\bibnamefont{Egelhof}},
  \bibinfo{author}{\bibfnamefont{H.}~\bibnamefont{Geissel}},
  \bibinfo{author}{\bibfnamefont{R.~E.} \bibnamefont{Grisenti}},
  \bibnamefont{et~al.}, \bibinfo{journal}{Nuclear Instruments and Methods in
  Physics Research B} \textbf{\bibinfo{volume}{317}}, \bibinfo{pages}{603}
  (\bibinfo{year}{2013}).

\bibitem[{\citenamefont{Steck and Litvinov}(2020)}]{SteckPPNP2020}
\bibinfo{author}{\bibfnamefont{M.}~\bibnamefont{Steck}} \bibnamefont{and}
  \bibinfo{author}{\bibfnamefont{Y.~A.} \bibnamefont{Litvinov}},
  \bibinfo{journal}{Progress in Particle and Nuclear Physics}
  \textbf{\bibinfo{volume}{115}}, \bibinfo{pages}{103811}
  (\bibinfo{year}{2020}).

\bibitem[{\citenamefont{Zhou et~al.}(2022)\citenamefont{Zhou, Yang, and the
  HIAF~project team}}]{ZhouAAPPS2022}
\bibinfo{author}{\bibfnamefont{X.}~\bibnamefont{Zhou}},
  \bibinfo{author}{\bibfnamefont{J.}~\bibnamefont{Yang}}, \bibnamefont{and}
  \bibinfo{author}{\bibnamefont{the HIAF~project team}},
  \bibinfo{journal}{AAPPS Bulletin} \textbf{\bibinfo{volume}{32}},
  \bibinfo{pages}{35} (\bibinfo{year}{2022}).

\bibitem[{\citenamefont{P\'alffy
  et~al.}(2007{\natexlab{b}})\citenamefont{P\'alffy, Evers, and
  Keitel}}]{PalffyPRL2007}
\bibinfo{author}{\bibfnamefont{A.}~\bibnamefont{P\'alffy}},
  \bibinfo{author}{\bibfnamefont{J.}~\bibnamefont{Evers}}, \bibnamefont{and}
  \bibinfo{author}{\bibfnamefont{C.~H.} \bibnamefont{Keitel}},
  \bibinfo{journal}{Phys. Rev. Lett.} \textbf{\bibinfo{volume}{99}},
  \bibinfo{pages}{172502} (\bibinfo{year}{2007}{\natexlab{b}}).

\bibitem[{\citenamefont{Fischer et~al.}(2019)\citenamefont{Fischer, Gaigalas,
  J\"onsson, and Biero\'n}}]{FischerCPC2019}
\bibinfo{author}{\bibfnamefont{C.~F.} \bibnamefont{Fischer}},
  \bibinfo{author}{\bibfnamefont{G.}~\bibnamefont{Gaigalas}},
  \bibinfo{author}{\bibfnamefont{P.}~\bibnamefont{J\"onsson}},
  \bibnamefont{and} \bibinfo{author}{\bibfnamefont{J.}~\bibnamefont{Biero\'n}},
  \bibinfo{journal}{Computer Physics Communications}
  \textbf{\bibinfo{volume}{237}}, \bibinfo{pages}{184} (\bibinfo{year}{2019}).

\bibitem[{\citenamefont{Minkov and P\'alffy}(2017)}]{MinkovPRL2017}
\bibinfo{author}{\bibfnamefont{N.}~\bibnamefont{Minkov}} \bibnamefont{and}
  \bibinfo{author}{\bibfnamefont{A.}~\bibnamefont{P\'alffy}},
  \bibinfo{journal}{Phys. Rev. Lett.} \textbf{\bibinfo{volume}{118}},
  \bibinfo{pages}{212501} (\bibinfo{year}{2017}).

\bibitem[{\citenamefont{Minkov and P\'alffy}(2019)}]{MinkovPRL2019}
\bibinfo{author}{\bibfnamefont{N.}~\bibnamefont{Minkov}} \bibnamefont{and}
  \bibinfo{author}{\bibfnamefont{A.}~\bibnamefont{P\'alffy}},
  \bibinfo{journal}{Phys. Rev. Lett.} \textbf{\bibinfo{volume}{122}},
  \bibinfo{pages}{162502} (\bibinfo{year}{2019}).

\bibitem[{\citenamefont{Minkov and P\'alffy}(2021)}]{MinkovPRC2021}
\bibinfo{author}{\bibfnamefont{N.}~\bibnamefont{Minkov}} \bibnamefont{and}
  \bibinfo{author}{\bibfnamefont{A.}~\bibnamefont{P\'alffy}},
  \bibinfo{journal}{Phys. Rev. C} \textbf{\bibinfo{volume}{103}},
  \bibinfo{pages}{014313} (\bibinfo{year}{2021}).

\bibitem[{\citenamefont{Shabaev et~al.}(2022)\citenamefont{Shabaev, Glazov,
  Ryzhkov, Brandau, Plunien, Quint, Volchkova, and Zinenko}}]{ShabaevPRL2022}
\bibinfo{author}{\bibfnamefont{V.~M.} \bibnamefont{Shabaev}},
  \bibinfo{author}{\bibfnamefont{D.~A.} \bibnamefont{Glazov}},
  \bibinfo{author}{\bibfnamefont{A.~M.} \bibnamefont{Ryzhkov}},
  \bibinfo{author}{\bibfnamefont{C.}~\bibnamefont{Brandau}},
  \bibinfo{author}{\bibfnamefont{G.}~\bibnamefont{Plunien}},
  \bibinfo{author}{\bibfnamefont{W.}~\bibnamefont{Quint}},
  \bibinfo{author}{\bibfnamefont{A.~M.} \bibnamefont{Volchkova}},
  \bibnamefont{and} \bibinfo{author}{\bibfnamefont{D.~V.}
  \bibnamefont{Zinenko}}, \bibinfo{journal}{Phys. Rev. Lett.}
  \textbf{\bibinfo{volume}{128}}, \bibinfo{pages}{043001}
  (\bibinfo{year}{2022}).

\bibitem[{\citenamefont{Schneider et~al.}(1991)\citenamefont{Schneider, Clark,
  Penetrante, McDonald, DeWitt, and Bardsley}}]{SchneiderPRA1991}
\bibinfo{author}{\bibfnamefont{D.}~\bibnamefont{Schneider}},
  \bibinfo{author}{\bibfnamefont{M.~W.} \bibnamefont{Clark}},
  \bibinfo{author}{\bibfnamefont{B.~M.} \bibnamefont{Penetrante}},
  \bibinfo{author}{\bibfnamefont{J.}~\bibnamefont{McDonald}},
  \bibinfo{author}{\bibfnamefont{D.}~\bibnamefont{DeWitt}}, \bibnamefont{and}
  \bibinfo{author}{\bibfnamefont{J.~N.} \bibnamefont{Bardsley}},
  \bibinfo{journal}{Phys. Rev. A} \textbf{\bibinfo{volume}{44}},
  \bibinfo{pages}{3119} (\bibinfo{year}{1991}).

\bibitem[{\citenamefont{Zhang et~al.}(2022)\citenamefont{Zhang, Wang, and
  Wang}}]{ZhangPRC2022}
\bibinfo{author}{\bibfnamefont{H.}~\bibnamefont{Zhang}},
  \bibinfo{author}{\bibfnamefont{W.}~\bibnamefont{Wang}}, \bibnamefont{and}
  \bibinfo{author}{\bibfnamefont{X.}~\bibnamefont{Wang}},
  \bibinfo{journal}{Phys. Rev. C} \textbf{\bibinfo{volume}{106}},
  \bibinfo{pages}{044604} (\bibinfo{year}{2022}).

\bibitem[{\citenamefont{Schneider et~al.}(1990)\citenamefont{Schneider, DeWitt,
  Clark, Schuch, Cocke, Schmieder, Reed, Chen, Marrs, Levine
  et~al.}}]{SchneiderPRA1990}
\bibinfo{author}{\bibfnamefont{D.}~\bibnamefont{Schneider}},
  \bibinfo{author}{\bibfnamefont{D.}~\bibnamefont{DeWitt}},
  \bibinfo{author}{\bibfnamefont{M.~W.} \bibnamefont{Clark}},
  \bibinfo{author}{\bibfnamefont{R.}~\bibnamefont{Schuch}},
  \bibinfo{author}{\bibfnamefont{C.~L.} \bibnamefont{Cocke}},
  \bibinfo{author}{\bibfnamefont{R.}~\bibnamefont{Schmieder}},
  \bibinfo{author}{\bibfnamefont{K.~J.} \bibnamefont{Reed}},
  \bibinfo{author}{\bibfnamefont{M.~H.} \bibnamefont{Chen}},
  \bibinfo{author}{\bibfnamefont{R.~E.} \bibnamefont{Marrs}},
  \bibinfo{author}{\bibfnamefont{M.}~\bibnamefont{Levine}},
  \bibnamefont{et~al.}, \bibinfo{journal}{Phys. Rev. A}
  \textbf{\bibinfo{volume}{42}}, \bibinfo{pages}{3889} (\bibinfo{year}{1990}).

\bibitem[{\citenamefont{Penetrante et~al.}(1991)\citenamefont{Penetrante,
  Bardsley, DeWitt, Clark, and Schneider}}]{PenetrantePRA1991}
\bibinfo{author}{\bibfnamefont{B.~M.} \bibnamefont{Penetrante}},
  \bibinfo{author}{\bibfnamefont{J.~N.} \bibnamefont{Bardsley}},
  \bibinfo{author}{\bibfnamefont{D.}~\bibnamefont{DeWitt}},
  \bibinfo{author}{\bibfnamefont{M.}~\bibnamefont{Clark}}, \bibnamefont{and}
  \bibinfo{author}{\bibfnamefont{D.}~\bibnamefont{Schneider}},
  \bibinfo{journal}{Phys. Rev. A} \textbf{\bibinfo{volume}{43}},
  \bibinfo{pages}{4861} (\bibinfo{year}{1991}).

\bibitem[{\citenamefont{Gu}(2008)}]{GuCJP2008}
\bibinfo{author}{\bibfnamefont{M.~F.} \bibnamefont{Gu}}, \bibinfo{journal}{Can.
  J. Phys.} \textbf{\bibinfo{volume}{86}}, \bibinfo{pages}{675}
  (\bibinfo{year}{2008}).

\bibitem[{fac()}]{facIaea2024}
\bibinfo{note}{FAC website at IAEA, {h}ttps://www-amdis.iaea.org/FAC/ (2024)}.

\bibitem[{\citenamefont{Kim and Pratt}(1983)}]{KimPRA1983}
\bibinfo{author}{\bibfnamefont{Y.~S.} \bibnamefont{Kim}} \bibnamefont{and}
  \bibinfo{author}{\bibfnamefont{R.~H.} \bibnamefont{Pratt}},
  \bibinfo{journal}{Phys. Rev. A} \textbf{\bibinfo{volume}{27}},
  \bibinfo{pages}{2913} (\bibinfo{year}{1983}).

\end{thebibliography}

\end{document}